\documentclass[preprint]{ptephy_v1}%%%%%% to generate preprint number
\usepackage{graphics} %for dealing with figures.
\usepackage{url}
\usepackage[dvipdfmx]{color}
\usepackage{bm}% bold math

\begin{document}

\title{Topological susceptibility of QCD with dynamical M\"obius domain wall fermions}

\newcommand{\Tsukuba}{
  Graduate School of Pure and Applied Sciences,
University of Tsukuba, Tsukuba, Ibaraki 305-8571, Japan
}

\newcommand{\CCS}{
Center for Computational Sciences,
University of Tsukuba, Ibaraki 305-8577, Japan
}
\newcommand{\KEK}{
  KEK Theory Center,
  High Energy Accelerator Research Organization (KEK),
  Tsukuba 305-0801, Japan
}
\newcommand{\GUAS}{
  School of High Energy Accelerator Science,
  The Graduate University for Advanced Studies (Sokendai),
  Tsukuba 305-0801, Japan
}
\newcommand{\YITP}{
  Center for Gravitational Physics, Yukawa Institute for Theoretical Physics, Kyoto 606-8502, JAPAN
}
\newcommand{\Osaka}{
  Department of Physics, Osaka University,
  Toyonaka 560-0043, Japan
}

\newcommand{\Edin}{
School of Physics and Astronomy, The University of Edinburgh, Edinburgh EH9 3JZ, United Kingdom
}

\author{S.~Aoki}
\affil{\YITP}
\author{G.~Cossu}
\affil{\Edin}
\author{H.~Fukaya}
\affil{\Osaka}
\author{S.~Hashimoto}
\author{T.~Kaneko}
\affil{\KEK}
\affil{\GUAS}
%\collaborator{JLQCD collaboration}

\begin{abstract}
We compute the topological susceptibility $\chi_t$ of lattice QCD
  with $2+1$ dynamical quark flavors described by
  the M\"obius domain wall fermion.
Violation of chiral symmetry as measured by the residual mass
is kept at $\sim$1 MeV or smaller.
We measure the fluctuation of the topological charge density
in a ``slab'' sub-volume of the simulated lattice using the method
proposed by Bietenholz {\it et al.}
The quark mass dependence of $\chi_t$
is consistent with the prediction of chiral perturbation theory, from which
the chiral condensate is extracted as
$\Sigma^{\overline{\rm MS}}(\mbox{2GeV}) = [274(13)(29)\mbox{MeV}]^3$,
where the first error is statistical and the second one is systematic.
Combining the results for the pion mass $M_\pi$ and decay constant $F_\pi$,
we obtain $\chi_t = 0.229(03)(13)M_\pi^2F_\pi^2$ at the physical point.
\end{abstract}
  %\begin{document}
\maketitle

\section{Introduction}

The topological susceptibility $\chi_t$ is an interesting quantity
which characterizes
how many topological excitations are created in the QCD vacuum.
Witten \cite{Witten:1979vv} and Veneziano \cite{Veneziano:1980xs} % Check name spells.
estimated $\chi_t$ in the large-$N_c$ (number of colors) limit 
and showed that it is proportional to the square of the $\eta^\prime$ meson mass.
In real QCD with $N_c=3$ and light dynamical quarks, however,
the argument of Witten and Veneziano is no longer valid.
It is not the $\eta^\prime$ meson but the (zero-momentum mode of) 
pion that controls the topological susceptibility.

According to the prediction of $SU(2)$ chiral perturbation theory (ChPT) at leading order (LO),
$\chi_t$ is expected to be proportional to the quark mass $m_{ud}$,
when the up and down quark masses are degenerate.
At one-loop, the quark mass dependence is predicted as \cite{Mao:2009sy, Aoki:2009mx, Guo:2015oxa}
\begin{equation}
\label{eq:ChPT}
\chi_t = \frac{m_{ud}\Sigma}{2}\left\{1-\frac{3m_{ud}\Sigma}{16\pi^2F_{\rm phys}^4}\ln \left(\frac{2m_{ud}\Sigma}{F_{\rm phys}^2M_{\rm phys}^2}\right)+\frac{4m_{ud}\Sigma}{F_{\rm phys}^4}l\right\},
\end{equation}
where $\Sigma$ denotes the chiral condensate, $l=l_3^r-l_7^r+h_1^r-h_3^r$ 
is a combination of the low-energy constants at next-to-leading order (NLO) \cite{Gasser:1983yg} 
and $M_{\rm phys}$(=135 MeV) and $F_{\rm phys}$(=92 MeV) are the physical values 
of the pion mass and decay constant, respectively.
Here $l_i^r$ are renormalized at $M_{\rm phys}$.
In the formula, we have assumed that the strange quark is decoupled
from the theory and $SU(2)$ chiral perturbation theory works.
In other words, the strange quark mass dependence is assumed to be
absorbed in the low energy constants.
%In the formula, we have assumed that the effect of the strange quark mass $m_s$ is negligible.
%In other words, these low energy constants are functions of $m_s$
%and evaluated at the physical strange quark mass.

%It is interesting to note that
By taking a ratio
with the ChPT predictions for the pion mass and decay constant (let us denote them by $M_\pi$ and $F_\pi$),
one can eliminate the chiral logarithm in ~(\ref{eq:ChPT}):
%%% Eq ratio %%%%%%%%%%%
\begin{eqnarray}
\label{eq:ratio}
\frac{\chi_t}{M_\pi^2F_\pi^2}=\frac{1}{4}\left[1+\frac{2M_\pi^2 l^{\prime}}{F_\pi^2}\right],
\end{eqnarray}
where $l^{\prime}=-l_4^r-l_7^r+h_1^r-h_3^r$ 
is again a combination of the NLO low energy constants,
which is independent of the renormalization scheme and scale at this order.
This ratio also cancels possible finite volume effects at NLO.
Moreover, the chiral limit of the ratio, 1/4, is protected from
the strange sea quark effects (see Appendix~\ref{app:3flavorChPT} for the details), 
as they always enter as a function of the ratio $m_{ud}/m_s$,
which can be absorbed into the (finite) renormalization of $l^{\prime}$.
We can therefore precisely estimate the topological susceptibility at the physical point
by measuring $\chi_t$, $M_\pi$, and $F_\pi$ at each simulation point.

It has been a challenging task for lattice QCD to compute $\chi_t$,
since it is sensitive to the discretization effects and
the violation of chiral symmetry \cite{Bazavov:2010xr,Cichy:2013rra, Bruno:2014ova} in particular.
This is partly because the quark mass dependence of $\chi_t$
is due to sea quarks, or a small quantum mechanical effect suppressed by $O(\hbar)$,
to which the discretization error is relatively large.
%Since it is a purely sea quark effect or a purely quantum effect of QCD,
%its sensitivity to the lattice spacing, 
%or violation of the chiral symmetry is known to be large .
%
%
%
%%% Ref. http://arxiv.org/pdf/1110.6013.pdf  ???
Even if we could simulate QCD on a sufficiently fine lattice,
the global topological charge would become frozen along the Monte Carlo history \cite{Schaefer:2010hu}.
Due to these difficulties, 
the study of the quark mass dependence 
and its comparison with the ChPT formula
of $\chi_t$ has been very limited, and only some pilot works with dynamical chiral 
fermions on rather small or coarse lattices 
\cite{Egri:2005cx, Aoki:2007pw, DeGrand:2007tx, Chiu:2008kt, Chiu:2008jq, Hsieh:2009zz, Aoki:2010dy, Chiu:2011dz, Chiu:2011za}
are available.

In this work, we improve the computation of $\chi_t$ in two ways.
One is to employ the domain wall fermion \cite{Kaplan:1992bt, Shamir:1993zy}
with an improvement by \cite{Brower:2004xi}, or known as the M\"obius domain wall fermion,
for the dynamical quarks, which enables us to precisely preserve chiral symmetry.
Even on our coarsest lattice,
the residual mass, related to the violation of the chiral symmetry, is kept at the order of $1$ MeV.
As will be shown below, our results show only a mild dependence of $\chi_t$ on
the lattice spacing, up to $a\sim 0.08$ fm.
The use of the domain-wall fermion allows us to sample configurations in different
topological sectors, which is also an advantage over
the simulation with the overlap fermion where we fixed the global topological charge in \cite{Aoki:2007pw}.

Another improvement comes from the use of sub-volumes of the simulated lattices.
Since the correlation length of QCD is limited, at most by $1/M_\pi$,
there is essentially no need to use the global topological charge to compute $\chi_t$.
The use of sub-volume was tested in our previous simulations with
overlap quarks \cite{Aoki:2007pw, Hsieh:2009zz} 
(see also \cite{Bautista:2015yza, Bietenholz:2016ymo}), where
the signal was extracted from finite volume effects, 
which have some sensitivity to $\chi_t$ \cite{Brower:2003yx, Aoki:2007ka}. 
%%% Slab method good.
In this work, we utilize a different method,
which was originally proposed by Bietenholz {\it et al.} \cite{Bietenholz:2015rsa, Bietenholz:2016szu}
(similar methods were proposed in \cite{Shuryak:1994rr} and \cite{deForcrand:1998ng}).
The method is based on a correlator, which gives a positive finite value 
even in the thermodynamical limit,
and thus is less noisy than our recent attempt in \cite{Fukaya:2014zda}\footnote{
See also Ref.~\cite{Brower:2014bqa} where a similar method to ours was attempted.
}.
We confirm that 30\%--50\% sub-volumes 
of the whole lattice, whose size is $\sim 2$ fm,  %%% CHECK !!! %%%%%%%%%%%%%%%%%%%%%%
are sufficient to extract $\chi_t$.
Moreover, the new definition shows more frequent fluctuation 
than that of the global topological charge on our finest lattice.

%Even on our finest lattice with the lattice spacing $a\sim 0.04$fm,
%the auto-correlation time is less than $200$ molecular-dynamics (MD) time,
%while the global topological charge is almost frozen during 
%the total $10000$ MD time we simulated.

We also employ a modern technique, the Yang-Mills (YM) gradient flow 
\cite{Luscher:2010iy, Luscher:2011bx, Bonati:2014tqa},
in order to make the global topological charge close to integers,
to remove the UV divergences, and to reduce the statistical noise.
With these improvements, we achieve good enough statistical precision to investigate
the dependence of $\chi_t$ on the sea quark mass.
In fact, our data of the topological susceptibility is consistent with the ChPT prediction~(\ref{eq:ChPT}),
from which the values of chiral condensate and $l^{\prime}$ are extracted.
%In fact, the topological susceptibility shows a good agreement with
%the ChPT prediction~(\ref{eq:ChPT}).
%We determine the value of chiral condensate, as well as $l^{\prime}$,
%taking the chiral and continuum limits from the formulas (\ref{eq:ChPT}),
%and (\ref{eq:ratio}).

The same set of data was also used to calculate the $\eta^\prime$ meson
mass \cite{Fukaya:2015ara}, 
which was extracted from the shorter distance region of 
the correlator of the topological charge density.
These two results show a nontrivial double-scale structure of 
topological fluctuation of the gauge field: it creates the $\eta^\prime$ 
meson at short distances, 
%while describing the vacuum mode of the pion at long distances.
while describing the vacuum mode of the pion (or the lowest mode,
which is constant over space-time) at long distances.

The rest of this paper is organized as follows.
First, we describe our lattice set-up in Sec.~\ref{sec:lattice-set-up}.
We then explain the method to extract the topological susceptibility
from the slab sub-volume in Sec.~\ref{sec:slabmethod}.
Our results at lower $\beta$ are presented in Sec.~\ref{sec:low-beta}.
Comparing the data with those obtained from global topology,
we examine the validity of our sub-volume method.
The results at higher $\beta$ are shown in Sec.~\ref{sec:high-beta},
and how we estimate the statistical errors is explained in
Sec.~\ref{sec:autocorrelation}.
Finally, we estimate the chiral and continuum limits
in Sec.~\ref{sec:finalresults} and give a summary in Sec.~\ref{sec:summary}.

%%%%%%%%%%%%%%%%%%%%%%%%%%%%%%%%%
\section{Lattice set-up}
\label{sec:lattice-set-up}
%%%%%%%%%%%%%%%%%%%%%%%%%%%%%%%%%

%\vspace*{5mm}%%% Lattice set-up
In the numerical simulation\footnote{
Numerical works are done with the QCD software package IroIro++ \cite{Cossu:2013ola}.
} of QCD,
we use the Symanzik gauge action and the M\"obius domain wall
fermion action for gauge ensemble generations \cite{Kaneko:2013jla, Noaki:2014ura, Cossu:2013ola}.
We apply three steps of stout smearing  of the gauge links \cite{Morningstar:2003gk}
for the computation of the Dirac operator.
Our main runs of $2+1$-flavor lattice QCD simulations are performed
with two different lattice sizes
$L^3\times T=32^3\times 64$ and $48^3\times 96$, for which
we set $\beta$ = 4.17 and 4.35, respectively.
The inverse lattice spacing $1/a$ is estimated to be 2.453(4)~GeV (for $\beta=4.17$) and 3.610(9)~GeV (for $\beta=4.35$),
using the input $\sqrt{t_0}=0.1465$ fm \cite{Borsanyi:2012zs}
where we use the reference YM gradient flow--time $t_0$, defined 
by $t^2\langle E\rangle |_{t=t_0}=0.3$ \cite{Luscher:2010iy} with the energy density $E$ of the gluon field.
The two lattices share a similar physical size $L\sim 2.6$ fm.
For the quark masses, we choose two values of 
the strange quark mass $m_s$ around its physical point, 
and 3--4 values of the up and down quark masses $m_{ud}$ for each $m_s$.
Since our data at the lightest pion mass around 230 MeV
($am_{ud}$ = 0.0035 at $\beta$ = 4.17)
may contain significant finite size effects,
we simulate a larger lattice $48^3\times 96$ with the same set of the parameters to check
if the finite volume systematics is small enough.
We also perform a simulation on a finer lattice 
$64^3\times 128$ (at $\beta=4.47$ [$1/a\sim 4.5$ GeV] and $M_\pi \sim$ 285 MeV).
For each ensemble, 500--1000 gauge configurations are sampled
from 5000 molecular dynamics (MD) time.
The ensembles used in this work are listed in Table.~\ref{tab:lattices}.

\begin{table}[h]
  \centering
  \begin{tabular}{|l|l|r|r|r|r|r|}
    \hline
    Lattice Spacing                    & $L^3\times T$                                 & $L_5$ & $a m_{ud}$ & $a m_s$ & $  m_\pi \text{ [MeV]} $ & $ m_{\pi}L $ \\
    \hline
    $\beta = 4.17,$                      & $ 32^3\times64$ $(L=2.6 \text{ fm})$ & 12    & 0.0035   & 0.040   & 230                    & 3.0          \\
    $a^{-1}=2.453(4)\text{ GeV}$    &                                                 &       & 0.0070   & 0.030   & 310                    & 4.0          \\
                                       &                                                 &       & 0.0070   & 0.040   & 310                    & 4.0          \\
                                       &                                                 &       & 0.0120   & 0.030   & 400                    & 5.2          \\
                                       &                                                 &       & 0.0120   & 0.040   & 400                    & 5.2          \\
                                       &                                                 &       & 0.0190   & 0.030   & 500                    & 6.5          \\
                                       &                                                 &       & 0.0190   & 0.040   & 500                    & 6.5          \\ \cline{2-7}
                                       & $48^3\times96 $ $ (L=3.9 \text{ fm})$             &  12     & 0.0035   & 0.040   & 230                    & 4.4          \\
    \hline
    $\beta= 4.35,$               & $48^3\times 96 $ $(L=2.6 \text{ fm})$           & 8     & 0.0042   & 0.018   & 300                    & 3.9          \\
    $a^{-1}=3.610(9)\text{ GeV}   $ &                                                 &       & 0.0042   & 0.025   & 300                    & 3.9          \\
                                       &                                                 &       & 0.0080   & 0.018   & 410                    & 5.4          \\
                                       &                                                 &       & 0.0080   & 0.025   & 410                    & 5.4          \\
                                       &                                                 &       & 0.0120   & 0.018   & 500                    & 6.6          \\
                                       &                                                 &       & 0.0120   & 0.025   & 500                    & 6.6          \\
    \hline
    $\beta = 4.47,$              & $64^3\times128 $ $(L=2.7  \text{ fm}) $          & 8     & 0.0030   & 0.015   & 280                    & 4.0          \\
    $a^{-1} = 4.496(9) \text{ GeV}$ &                                                 &       &          &         &                        &              \\
    \hline
  \end{tabular}
  \caption{Parameters of the JLQCD gauge ensembles used in this work. Pion masses are rounded to the nearest $10$ MeV. 
%The ensemble with $m_\pi L \approx 3.0$ is excluded in all analysis below to avoid possible finite volume effects.  
\label{tab:lattices}}
\end{table}
%%%% FLAG Mpi L should be used ???????%%%%%%%%%%%%%%%%%%%%%%%%%%%%%%%%%%%%%%%%%%

In this setup, we confirm that the violation of the chiral symmetry in
the M\"obius domain wall fermion formalism
is well under control.
The residual mass is $\sim 1$~MeV \cite{Hashimoto:2014gta}
by choosing the lattice size in the fifth direction $L_5$ = 12 at $\beta$ = 4.17
and less than 0.2 MeV with $L_5$ =  8  at $\beta$ = 4.35 (and 4.47).

On generated configurations, we perform 500--1640 steps of the
YM gradient flow (using the conventional Wilson gauge action) 
with a step-size $\Delta t/a^2=$0.01. 
At every 200--400 steps (depending on the parameters)
we store the configuration of the topological charge density.
The two-point correlator is measured using the Fast Fourier Transform (FFT) technique
to average source and sink points over whole lattice sites.

In the following analysis, we measure the 
integrated auto-correlation time of every quantity, 
following the method proposed in \cite{Luscher:2005rx, Schaefer:2010hu}.
The statistical error is estimated by the jackknife method (without binning)
multiplied by the square root of auto-correlation time 
normalized by the MD time interval of the configuration samples.
We will discuss more details about the auto-correlation time of topological fluctuations in Sec.~\ref{sec:autocorrelation}.

The pion mass and decay constant are computed combining
the pseudoscalar correlators with local and smeared source operators.
Details of the computation are presented in a separate article \cite{Fahy:2015xka}.
%{\bf Needs here a table of mres}
%{\bf Results table mpi fpi tau tau <Q> <Q2> chit}

%%%%%%%%%%%%%%%%%%%%%%%%%%%%%%%%%%%%%%%%%%%%%%%%
\section{Topological susceptibility in a ``slab''}
\label{sec:slabmethod}
%%%%%%%%%%%%%%%%%%%%%%%%%%%%%%%%%%%%%%%%%%%%%%%%

\begin{figure*}[tbp]
  \centering
  \includegraphics[width=7cm]{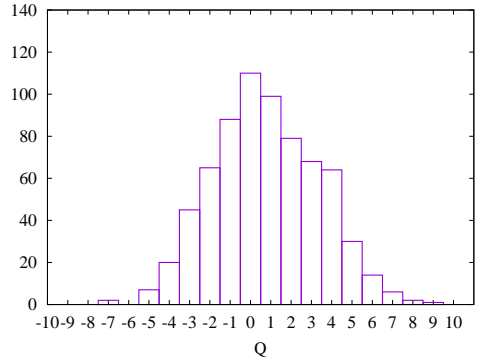}
  \caption{
    The distribution of $\sum_x q^{\rm lat}(x)$ at $\beta=4.17$, $m_{ud}=0.007$, and $m_s=0.04$.
  }
\label{fig:Qdist}
\end{figure*}

%\vspace*{5mm}
We use the conventional gluonic definition of the topological charge density
$q^{\rm lat}(x)$, the so-called clover construction \cite{Bruno:2014ova}. 
Since the YM gradient flow smooths the gauge field
in the range of $\sqrt{8t}\sim 0.5$ fm of the lattice,
a simple summation $Q_{\rm lat}=\sum_x q^{\rm lat}(x)$ over the whole sites gives
values close to integers, as shown in Fig.~\ref{fig:Qdist}. 
%Since deviation from integers is quite linear,
%we may define a (finite) renormalization factor
%\begin{equation}
%Z_q = \frac{Q_{\rm lat}}{\mbox{round}(Q_{\rm lat})},
%\end{equation}
%determined by a linear fit to the lattice data.
%Here ``round'' means an integer of rounded $Q_{\rm lat}$.
%We confirm that $Z_q$ approaches to unity at high $\beta$.
%For example, at the simulated pion mass $M_\pi\sim 300$ MeV,
%$Z_q=0.911(6)$, 0.968(7), 0.983(3) at $\beta=$4.17, 4.35, 4.47, respectively.
%We also confirm that the quark mass dependence of $Z_q$ is negligibly small.
%However, we find that their squared $\langle Q_{\rm lat}^2 \rangle$ without
%multiplying $1/Z_q^2$ is
%consistent with $\langle \mbox{round}(Q_{\rm lat})^2 \rangle$.
%This is probably because another lattice artifact due to nonzero contribution from
%would-be zero-instanton sectors, partly cancels the effect of $Z_q$.
%Therefore, we use the bare operator $q^{\rm lat}(x)$ in the following analysis.

\begin{figure*}[bthp]
  \centering
  \includegraphics[width=9.5cm]{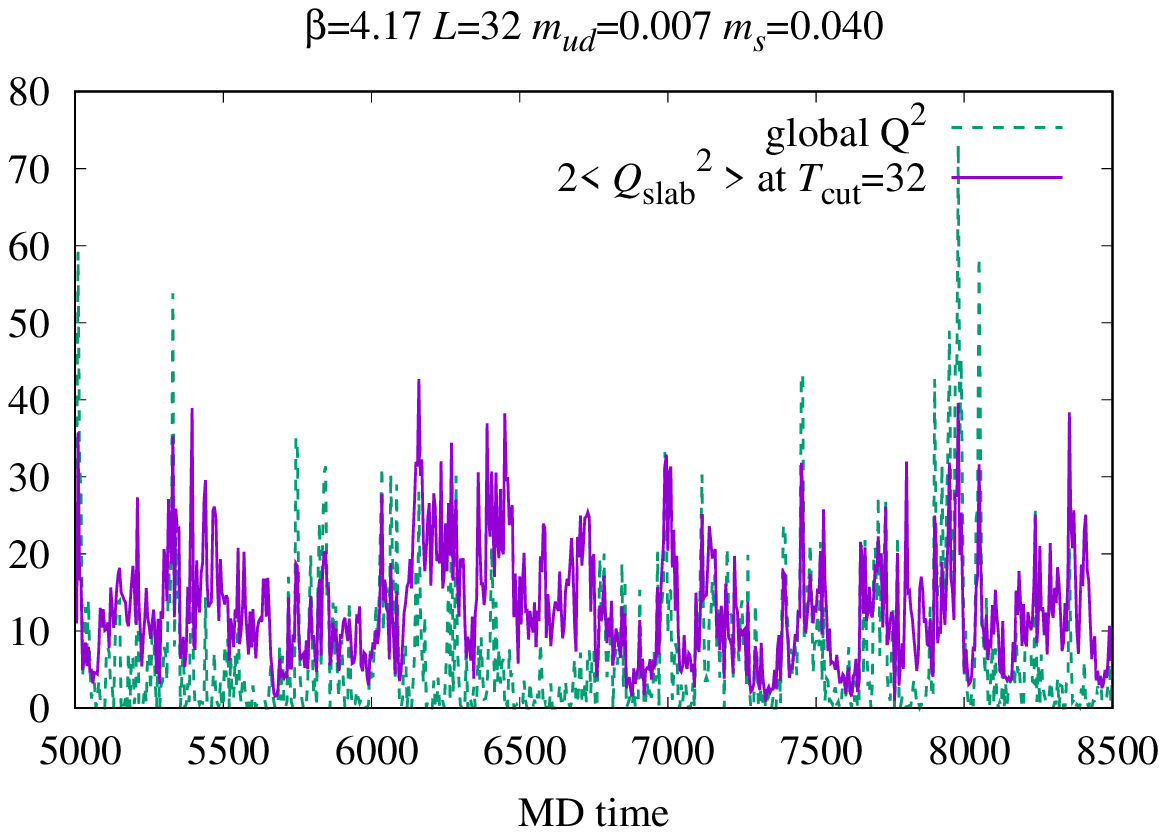}
 \includegraphics[width=9.5cm]{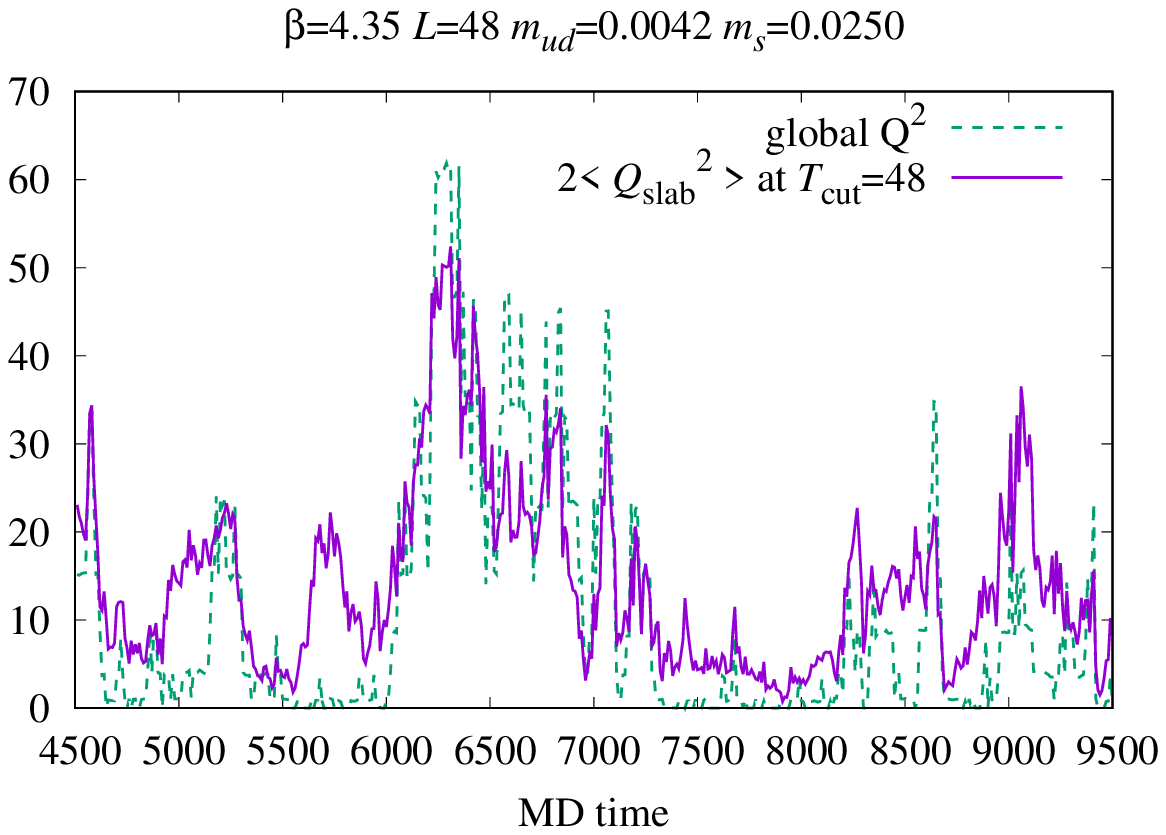}
  \includegraphics[width=9.5cm]{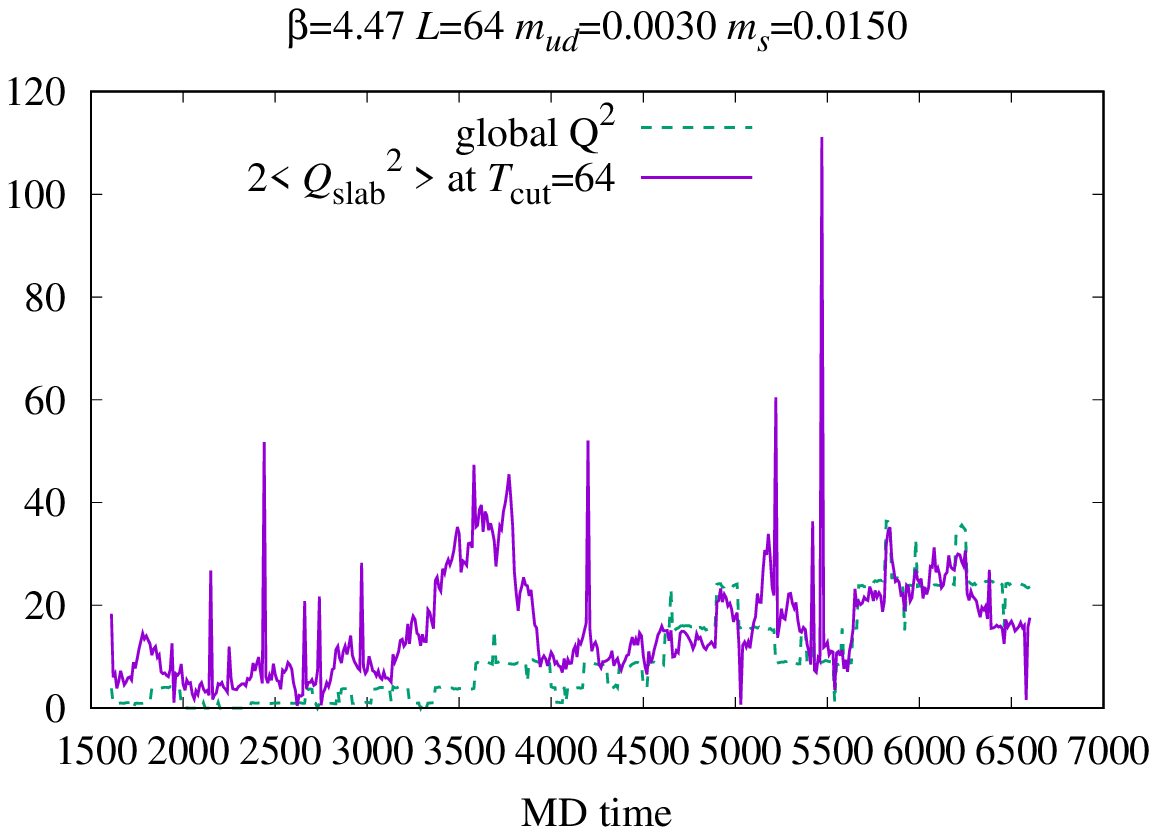}
\caption{
  MD history of $\langle Q_{\rm slab}^2(T_{\rm cut}=T/2)\rangle$ (solid lines)
  and that of global topological charge $Q^2$ (dashed).
  Data at $\beta=4.17, m_{ud}=0.007,m_s=0.040$ (top panel) and those at
  $\beta=4.35, m_{ud}=0.0042,m_s=0.0250$ (middle) and at
  $\beta=4.47, m_{ud}=0.0030, m_s=0.0150$ (bottom) are shown.
  These three simulations share a similar value of the pion mass $\sim 300$ MeV and physical volume.
}
\label{fig:autocorrelation}
\end{figure*}

As is well known, the global topological charge $Q_{\rm lat}$ 
suffers from long auto-correlation time in lattice simulations, especially when the lattice spacing is small.
This is true also in our simulations, 
as shown in Fig.~\ref{fig:autocorrelation}.
At the highest $\beta=4.47$, $Q_{\rm lat}$ drifts very slowly with auto-correlation time of possibly $O(1000)$.
It is, therefore, not feasible to estimate $\chi_t$ without performing much longer runs.
The details of the auto-correlation time of the topological charge and
its density operator will be discussed in Section~\ref{sec:autocorrelation}.

%as shown in the left panel of Fig.~\ref{fig:Qhistory}.

%\begin{figure*}[tbp]
%  \centering
%  \includegraphics[width=7cm]{Tophistory_beta4.47.eps}
%  \includegraphics[width=7cm]{history_chit_bulk_beta4.47.eps}
%  \caption{
%Monte Carlo history of the global topological charge $Q$ (left) and $\chi_t^{\rm sub V}$ (right).
%Data at our highest $\beta=4.47$ (with $m_{ud}=0.003$ and $m_s=0.015$) are shown.
%  }
%\label{fig:Qhistory}
%\end{figure*}

Instead of using the global topological charge $Q_{\rm lat}$, 
we attempt to extract the topological susceptibility from 
a sub-volume $V_{\rm sub}$ of the whole lattice $V$.
Since the correlation length of QCD is limited by at most $1/M_\pi$,
the subvolume $V_{\rm sub}$ should contain sufficient information
to extract $\chi_t$, provided that its size is larger than $1/M_\pi$.
%there is no theoretical need to use the global topological 
%one may compute $\chi_t$ within $V_{\rm sub}$ whose size is longer than $1/M_\pi$.
One can then effectively increase the statistics by
$V/V_{\rm sub}$, since each piece of $V/V_{\rm sub}$ sub-lattices
may be considered as an uncorrelated sample.
Moreover, there is no potential barrier among topological sectors:
the instantons and anti-instantons freely come in and go out of
the sub-volume, which should make the auto-correlation time
of the observable shorter than that of the global topological charge.

There are various ways of cutting the whole lattice into sub-volumes
and compute the correlation functions in them.
After some trials and errors, we find that the so-called
``slab'' method, proposed by Bietenholz {\it et al.} \cite{Bietenholz:2015rsa}
is efficient for the purpose of computing $\chi_t$.
The idea is to sum up the two-point correlators of the topological charge density,
over $x$ and $y$ in the same sub-volume:
\begin{eqnarray}
%\chi_t^{\rm subV} &\equiv& \frac{V}{V-V_{sub}}\int_{V_{sub}} d^4x 
%\left\langle \left(q(x)-\frac{Q}{V}\right)\left(q(y)-\frac{Q}{V}\right)\right\rangle,\\
\langle Q_{\rm slab}^2(T_{\rm cut}) \rangle &\equiv& \int_{T_{\rm ref}}^{T_{\rm cut}+T_{\rm ref}} 
dx_0 \int_{T_{\rm ref}}^{T_{\rm cut}+T_{\rm ref}} 
dy_0\int d^3x \int d^3y\left\langle q^{\rm lat}(x)q^{\rm lat}(y)\right\rangle.
\end{eqnarray}
Here the integration over $x$ and $y$ in the spatial directions runs in the whole spatial volume
(since the YM gradient flow is already performed, there is no divergence from the points of $x=y$), 
while the temporal sum is restricted to the region 
$[T_{\rm ref}, T_{\rm cut}+T_{\rm ref}]$, which is called a ``slab''.
Here $T_{\rm ref}$ is an arbitrary reference time.
Due to the translational invariance, the slabs
sharing the same thickness $T_{\rm cut}$
are physically equivalent, and one can average over $T_{\rm ref}$.
This method is statistically more stable than the other sub-volume method we applied in
\cite{Aoki:2007pw, Hsieh:2009zz}
because $\langle Q_{\rm slab}^2(T_{\rm cut}) \rangle$ is guaranteed to be always positive.

If we sample large statistics on a large enough lattice volume,
$\langle Q_{\rm slab}^2(T_{\rm cut}) \rangle$ should be just $T_{\rm cut}/T$
of $\chi_t V$. Namely $\langle Q_{\rm slab}^2(T_{\rm cut}) \rangle$ 
should be a linear function in $T_{\rm cut}$.
Its leading finite volume correction can be estimated
using the formula in \cite{Aoki:2007ka}:
\begin{eqnarray}
\label{eq:linear}
\langle Q_{\rm slab}^2(T_{\rm cut}) \rangle = (\chi_t V)\times \frac{T_{\rm cut}}{T}
+C(1-e^{-m_0 T_{\rm cut}})(1-e^{-m_0 (T-T_{\rm cut})}),
\end{eqnarray}
where $C$ is an unknown constant, and $m_0$ is the mass of the 
first excited state, the $\eta'$ meson\footnote{
The finite volume effects are due to propagation of the
mesons in the flavor singlet channel.
As the ground state or the $\eta'$ meson is heavy,
we neglect the higher order effects.
Even if we include them, the structure of linear+constant
in Eq.~\ref{eq:linear} is unchanged since
their effect is just an additional constant.
}.
Note that for $1/m_0 \ll T_{\rm cut}\ll T-1/m_0$, the formula gives a simple 
linear function in $T_{\rm cut}$ plus a constant.
Also, note that in the limit of $T_{\rm cut}=T$, 
$\langle Q_{\rm slab}^2(T_{\rm cut}=T) \rangle = \langle Q^2\rangle = \chi_t V$.

%For a fixed topological sector of global topological charge $Q$, 
%we obtain a modified formula:
%\begin{eqnarray}
%\langle Q_{\rm slab}^2(T_{\rm cut}) \rangle_Q = (\chi_t V)\times \frac{T_{\rm cut}}{T}
%+\frac{T_{\rm cut}^2}{T^2}\left(Q^2-\chi_t V\right)
%+C(1-e^{-m_0 T_{\rm cut}})(1-e^{-m_0 (T-T_{\rm cut})}),
%\end{eqnarray}
%upto power corrections in $1/V$. In this case, the formula is
%a quadratic function in $T_{\rm cut}$.

%Using the linear or quadratic behaviors, we can extract the
%topological susceptibility from the slab.

Assuming the linearity in $T_{\rm cut}$, one
can extract the topological susceptibility through
%%%  chit definition %%%%%%%%%%%%%%
\begin{eqnarray}
\label{eq:chitslab1}
\chi_t^{\rm slab} = \frac{T}{V}\left[\frac{\langle Q_{\rm slab}^2(t_1) \rangle - \langle Q_{\rm slab}^2(t_2) \rangle}{t_1-t_2}\right],
\end{eqnarray}
with two reference thicknesses $t_1$ and $t_2$.
In our numerical analysis, $T_{\rm ref}$ is averaged over the temporal direction.
Since the data at $t_i$ and $T-t_i$ are not independent,
we choose $t_1$ and $t_2$ in a range 1.6 fm $<t_1, t_2< T/2$. 
In the numerical analysis,
we replace $q^{\rm lat}(x)$ by $q^{\rm lat}(x)-\langle Q_{\rm lat}/V\rangle$
to cancel a possible bias due to the long auto-correlation of the global topology.

 The original proposal in \cite{Bietenholz:2015rsa} mainly used
  the correlator in a fixed topological sector.
  The formula corresponding to (\ref{eq:linear}) then contains a subtraction of the
  contribution from the global topology. We find that the statistical noise is larger
  with this choice while the results from different topological sectors are consistent.
  In the following analysis, we use (\ref{eq:linear}) after summing over the topological sectors.

We find that the signal using this slab method
is less noisy than the previous attempts
in \cite{Aoki:2007pw, Hsieh:2009zz}.
Moreover, as shown in Fig.~\ref{fig:autocorrelation} 
and will be discussed in details later,
the new definition shows more frequent fluctuation 
than that of the global topological charge on our finest lattice.

%  However, we find that the fixed topology formula subtracts contribution
%  from the global topology, which enhances the statistical errors.
%  We, therefore, use the formula after summing all topological sectors in this work.

%%%%%%%%%%%%%%%%%%%%%%%%%%%%%%%%%%%%%%%%%%%%%%%%
\section{Results at low $\beta$}
\label{sec:low-beta}
%%%%%%%%%%%%%%%%%%%%%%%%%%%%%%%%%%%%%%%%%%%%%%%%

At $\beta=4.17$, which corresponds to the lattice spacing $a\sim 0.08$ fm,
both the global topological charge $Q_{\rm lat}$ and $Q_{\rm slab}^2(T_{\rm cut})$
fluctuate well, as shown in the top panel of Fig.~\ref{fig:autocorrelation}.
The data on this lattice, therefore, provide a good testing ground to 
examine the validity of the slab sub-volume method, comparing 
with the naive definition of the topological susceptibility with $\langle Q_{\rm lat}^2\rangle/V$.

\begin{figure*}[tpbh]
  \centering
 \includegraphics[width=10cm]{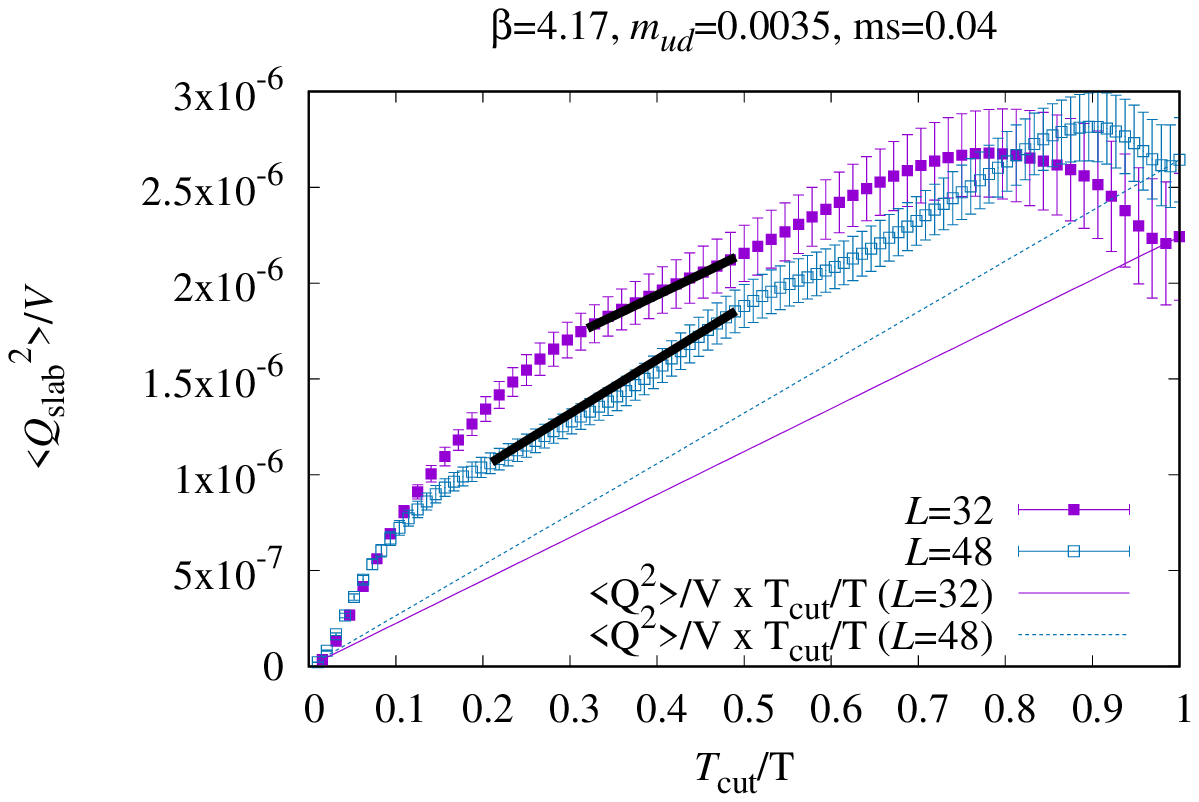}
\caption{
 $\langle Q_{\rm slab}^2(T_{\rm cut})\rangle$ as a function of $T_{\rm cut}/T$.
  Data at lightest mass $m_{ud}=0.0035$, $\beta=4.17$ with two different lattice sizes
  $L=32$ and $L=48$ are shown. $T=2L$ for both lattices.
  The solid and dotted lines show the slope obtained from the global topological charge
  measured on the $L=32$ and $L=48$ lattices, respectively.
 Two end-points of the thick line segments show the reference points $t_1$ and $t_2$
 taken for determination of the topological susceptibility.
 Note that the value of $t_1=20$ is the same for the two data.
% They show a quick convergence to the linear + constant function
% given in ~(\ref{eq:linear}) already at $T_{\rm cut}=20\sim 1.6$fm.
% The slope is consistent with that from global topology 
% shown in solid and dashed lines. The slopes of $L=48$ and $L=32$ results show 
% a good agreement within the statistical errors.
}
\label{fig:finiteV2}
\end{figure*}

In Fig.~\ref{fig:finiteV2}, $\langle Q_{\rm slab}^2(t_{\rm cut}) \rangle$
observed at the lightest sea quark mass $m_{ud}=0.0035$, $\beta=4.17$
on two different volumes $L=32$ and $L=48$ are plotted as a function of $T_{\rm cut}/T$.
 The data  converge to a linear plus constant function
 given in ~(\ref{eq:linear}) at $T_{\rm cut}=20$, which corresponds to $\sim 1.6$ fm.
 The slope, or $\chi_t^{\rm slab}$, is consistent with that from global topology 
 shown by solid and dotted lines for $L=32$ and $L=48$ lattices, respectively.  
 We also observe the consistency between the $L=32$ and $L=48$ data,
 which suggests that the systematics due to the finite volume is well under control.

\begin{figure*}[bthp]
  \centering
 \includegraphics[width=10cm]{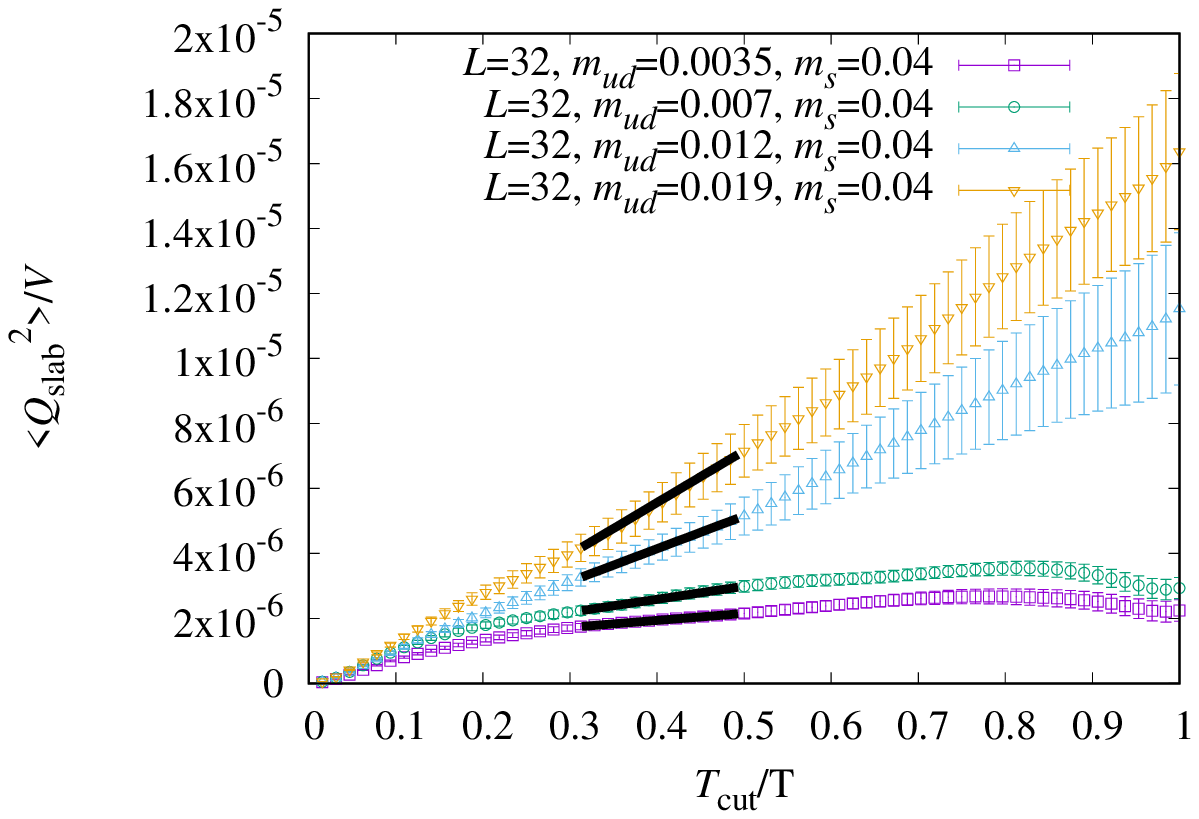}
\includegraphics[width=10cm]{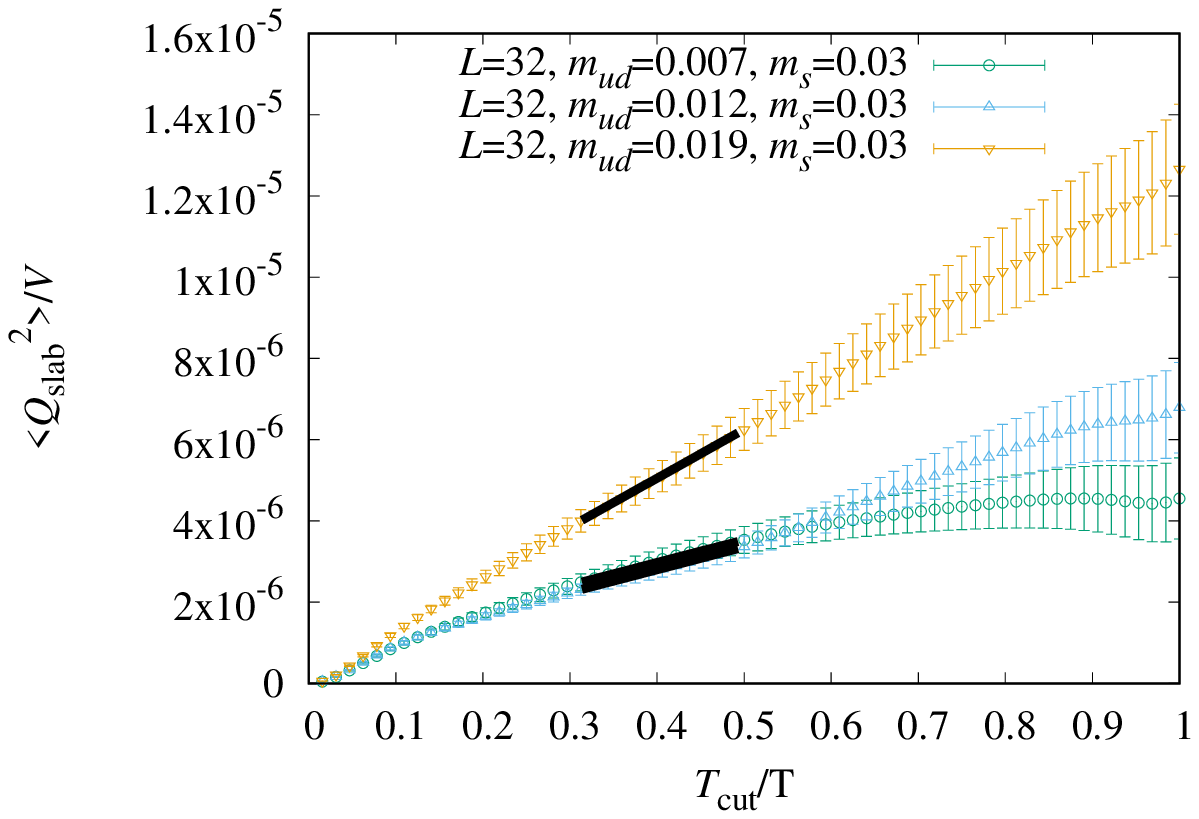}
\caption{
  Same as Fig.~\ref{fig:finiteV2} but at different up and down quark masses.
  Data at $\beta=4.17$ and $m_s=0.04$ (top panel) and those at $m_s=0.03$ (bottom) are shown.
}
\label{fig:chitb4.17}
\end{figure*}

The ``linear $+$ constant'' behavior is also seen in ensembles with heavier quark masses,
as presented in Fig.~\ref{fig:chitb4.17}.

\begin{figure*}[bthp]
  \centering
 \includegraphics[width=10cm]{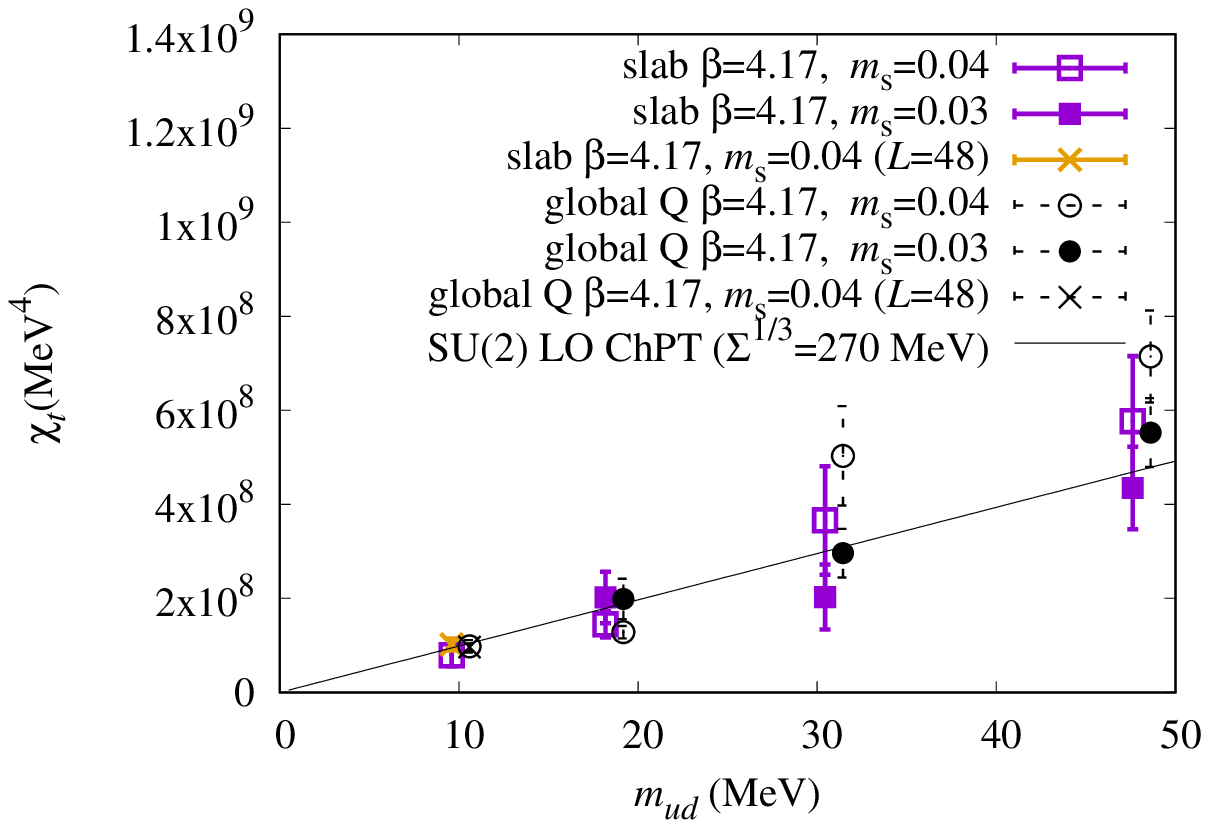}
\caption{
  $m_{ud}$ dependence of topological susceptibility at $\beta=4.17$ 
  obtained from $\langle Q_{\rm slab}^2(T_{\rm cut})\rangle$ (solid symbols)
  and those from the global topological charge
  (dashed, slightly shifted to avoid overlapping with the former data).
  The LO prediction from $SU(2)$ ChPT,
  where the chiral condensate $\Sigma^{1/3}=270$ MeV,  is also shown
  for an eye-guide.
}
\label{fig:chiral4.17}
\end{figure*}

The extracted values of the topological susceptibility
from the slope,
show a good agreement with the ChPT prediction,
as shown in Fig.~\ref{fig:chiral4.17} by open and filled squares.
The leading-order ChPT formula, $\chi_t = m_{ud}\Sigma/2$,
with $\Sigma =[270 \mbox{MeV}]^3$ (solid line) is drawn as an eye guide.
In the same plot, we also plotted the estimate for $\chi_t$ obtained from
the global topological charge by circles, which again agree with the results,
validating the slab method.
The values of $\chi_t^{\rm slab}$ are listed in Table~\ref{tab:results1}.
How we estimate their errorbars is explained in the following two sections.


%%%%%%%%%%%%%%%%%%%%%%%%%%%%%%%%%%%%%%%%%%%%%%%%
\section{Results at high beta}
\label{sec:high-beta}
%%%%%%%%%%%%%%%%%%%%%%%%%%%%%%%%%%%%%%%%%%%%%%%%

\begin{figure*}[bthp]
  \centering
 \includegraphics[width=10cm]{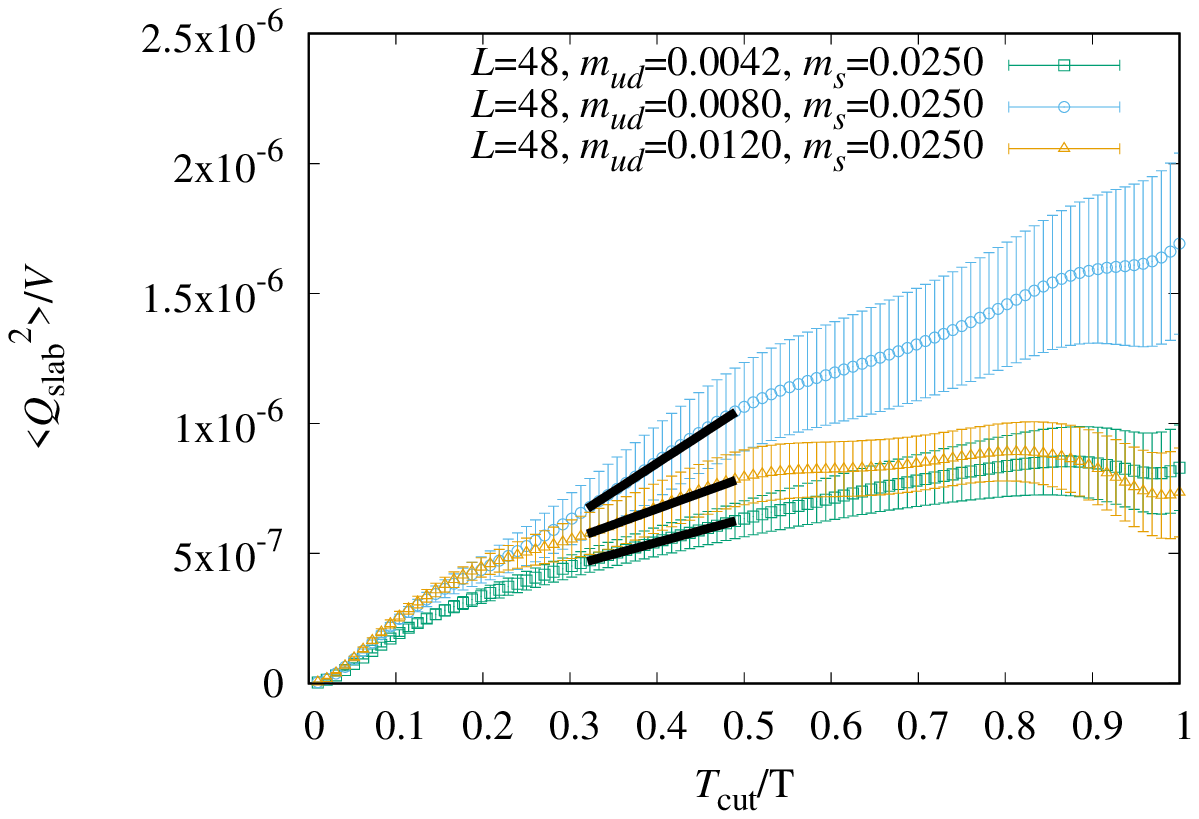}
 \includegraphics[width=10cm]{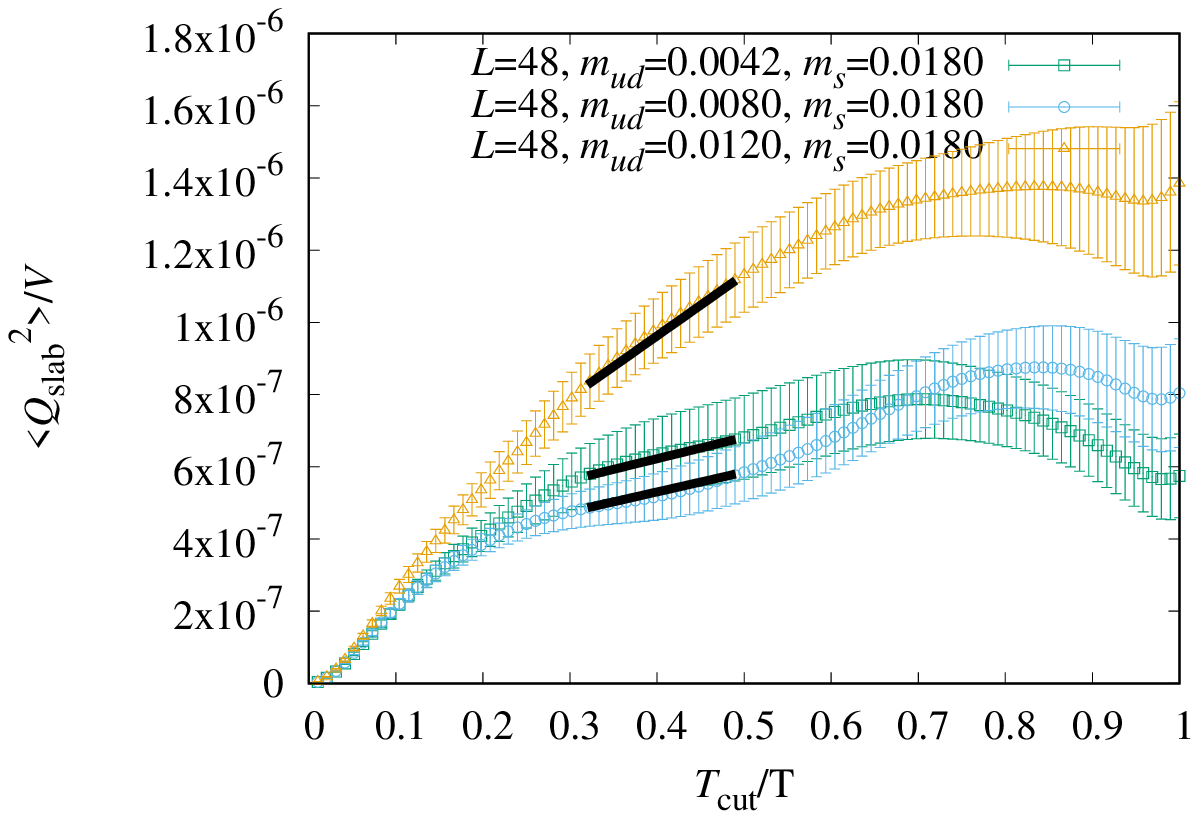}
  \includegraphics[width=10cm]{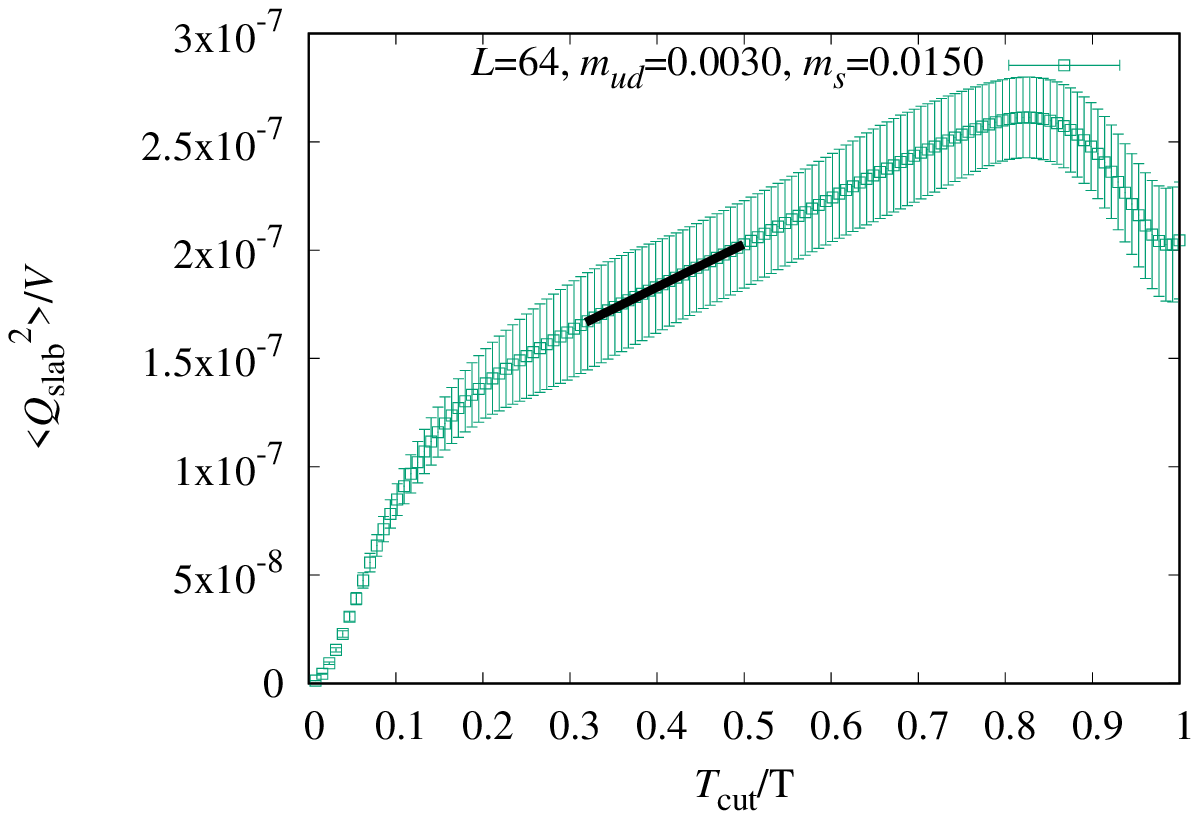}
\caption{
  $\langle Q_{\rm slab}^2(T_{\rm cut})\rangle$ at different up and down quark masses.
  Data at $\beta=4.35$ and $m_s=0.0180$ (top) and $m_s=0.0250$ (middle) 
  and those at $\beta=4.47$ and $m_s=0.0150$ (bottom) are shown.
}
\label{fig:chithighbeta}
\end{figure*}

At higher $\beta$ values, we still find a reasonable
slope at the lightest quark mass for each $\beta$ and $m_s$,
as shown in Fig.~\ref{fig:chithighbeta}.
For heavier masses, however, some curvature is seen.
We consider this curvature is an effect from the bias of the global topological charge.
This observation is consistent with previous works (see \cite{Schaefer:2010hu} for example),
which reported that the heavier pion mass ensembles
show the longer auto-correlation of the topological charge,
and the larger deviation of $\langle Q_{\rm lat}\rangle$ from zero.
We determine the reference $t_1\sim 1.6$ fm
using data at the lightest quark mass and always choose $t_2=T/2 \sim 2.6$ fm.
In order to estimate the systematic errors due to non-linear behavior,
we compare the results 
with 1) those obtained from different reference times $(t_1', t_2')=(t_1,\frac{t_1+t_2}{2})$, and $(\frac{t_1+t_2}{2},t_2)$,
and 2) those obtained without the subtraction of $\langle Q\rangle/V$ in the definition of the topological charge density.
The larger deviation is treated as a systematic error.
More details are presented in Appendix~\ref{app:bias}.

\begin{figure*}[bthp]
  \centering
 \includegraphics[width=10cm]{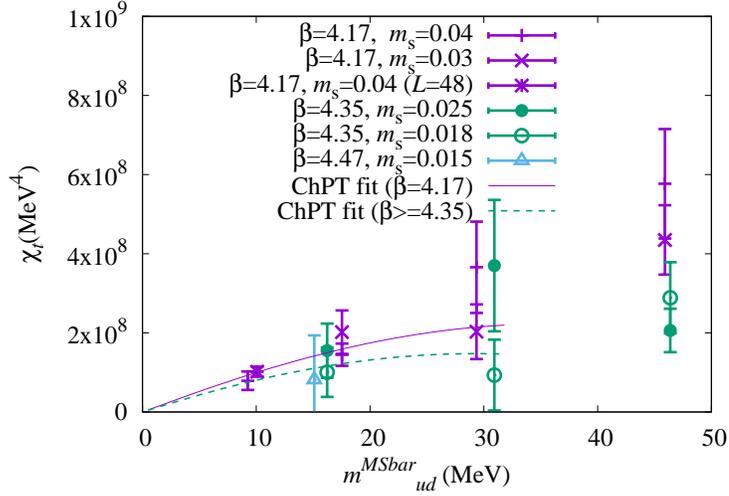}
\caption{
 $m_{ud}$ dependence of topological susceptibility obtained from
  the slab sub-volume method.
  The heaviest four points are not included in the fit.
}
\label{fig:result}
\end{figure*}

\begin{figure*}[bthp]
  \centering
 \includegraphics[width=10cm]{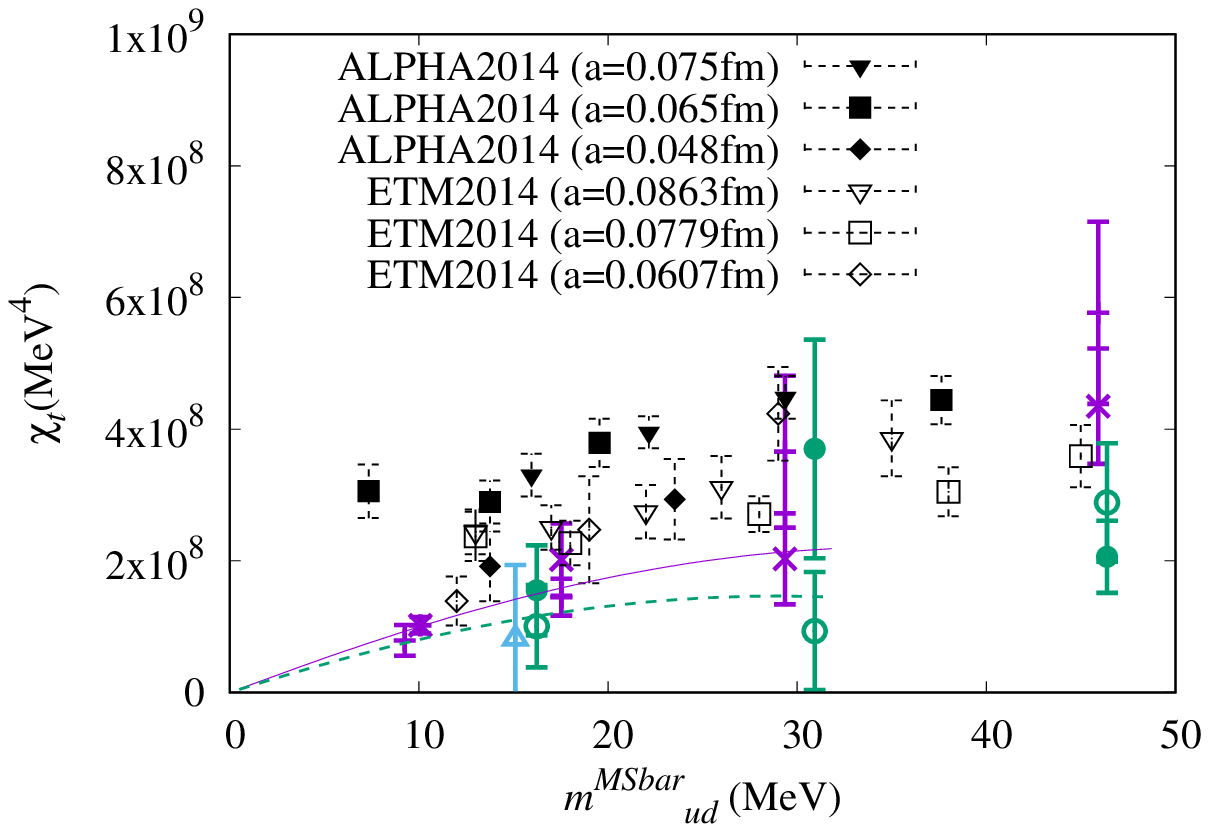}
 \caption{
   The same figure as Fig.~\ref{fig:result} but a comparison with
   Ref.~\cite{Cichy:2013rra} (ETM2014, $N_f=2+1+1$ results converted using the input $r_0=0.46$ fm)
   and  Ref.~\cite{Bruno:2014ova}
   (ALPHA2014, $N_f=2$ results converted assuming $m_{ud}=M_\pi^2F_\pi^2/(2 \Sigma)$
   using the inputs $t_1=0.061$ fm$^2$, $F_\pi=92$ MeV and $\Sigma=$(270 MeV)$^3$)
   is shown.
 }
\label{fig:result-comparison}
\end{figure*}

Our results are summarized in Fig.~\ref{fig:result}
(see also Fig.~\ref{fig:result-comparison} for a comparison
with Ref.~\cite{Cichy:2013rra} and  Ref.~\cite{Bruno:2014ova}).
Although the data at higher $\beta$ are rather scattered
compared to those at $\beta=4.17$, they can be used to estimate the chiral condensate $\Sigma$,
  assuming the linear suppression around the chiral limit.
%the linear suppression
%around the chiral limit can still be observed.
%They are good enough to estimate the chiral condensate $\Sigma$,
%comparing with the ChPT prediction.
Before going to the details,
we discuss the auto-correlation of $\chi_t^{\rm slab}$ and
show how we estimate the statistical errors in the next section.


%%%%%%%%%%%%%%%%%%%%%%%%%%%%%%%%%%%%%%%%%%%%%%%%
\section{Auto-correlation and error estimates}
\label{sec:autocorrelation}
%%%%%%%%%%%%%%%%%%%%%%%%%%%%%%%%%%%%%%%%%%%%%%%%

Gauge configurations generated by a Markov chain
are generally not independent but have auto-correlations.
How much they are correlated depends on observables.
We therefore need to carefully measure the
auto-correlation of the target observable $O$:
%%%  auto-correlation
\begin{eqnarray}
\Gamma_O(\Delta \tau)=\langle O(\tau)O(\tau+\Delta \tau)\rangle_\tau,
\end{eqnarray}
where $\tau$ denotes the Monte Carlo time,
and the average $\langle\cdots \rangle_{\tau}$ is taken over $\tau$.

When the Monte Carlo trajectory is long enough,
compared to the auto-correlation time of any observables,
one can estimate the so-called integrated auto-correlation time by
%%% Eq tau int %%%%%%%%%%
\begin{eqnarray}
  \label{eq:tauint}
  \tau_{\rm int} = \frac{1}{2} +
  \sum_{\Delta \tau =0}^W \rho(\Delta \tau),\;\;\;
   \rho(\Delta \tau)=\frac{\Gamma_O(\Delta \tau)}{\Gamma_O(0)},
\end{eqnarray}
where the upper end of the summation window $W$ is chosen to
where $\rho(\Delta \tau)$ becomes consistent with zero within
the error.
The above formula assumes that $\Gamma_O(\Delta \tau)$
converges to a single exponential function well below $W$.

If the observables suffer from long auto-correlation,
and the Monte Carlo trajectory is not long enough, on the other hand,
the above procedure may underestimate the auto-correlation time,
since some very slow decay modes can be hidden in the
error of $\rho(\Delta \tau)$.
This problem is similar to that of hadron spectroscopy with a short temporal  extension,
where one does not have long enough fitting range to disentangle
the ground state from excited states, which leads to over-estimation of the mass.

The ALPHA collaboration \cite{Schaefer:2010hu}
carefully studied the effect of slow modes,
and proposed an improved estimate of the auto-correlation time,
%%% Eq tau improved %%%%%%%
\begin{eqnarray}
\tau_{\rm imp} = \tau_{\rm int}' + \tau_{\rm exp}\rho(W'),
\end{eqnarray}
where $\tau_{\rm int}'$ is the same summation as (\ref{eq:tauint})
but with a smaller upper bound $W'$ where
$\rho(W')$ becomes lower than 3/2  standard deviations. 
$\tau_{\rm exp}$ is the auto-correlation of the slowest mode.
The proposal is equivalent to considering
a continuation of $\Gamma_O(\Delta \tau)$ at $\Delta \tau = W'$
to the slowest possible exponential function $\Gamma_O(W') \exp(-(\Delta \tau-W')/ \tau_{\rm exp})$.

\begin{figure*}[bthp]
  \centering
 \includegraphics[width=8cm]{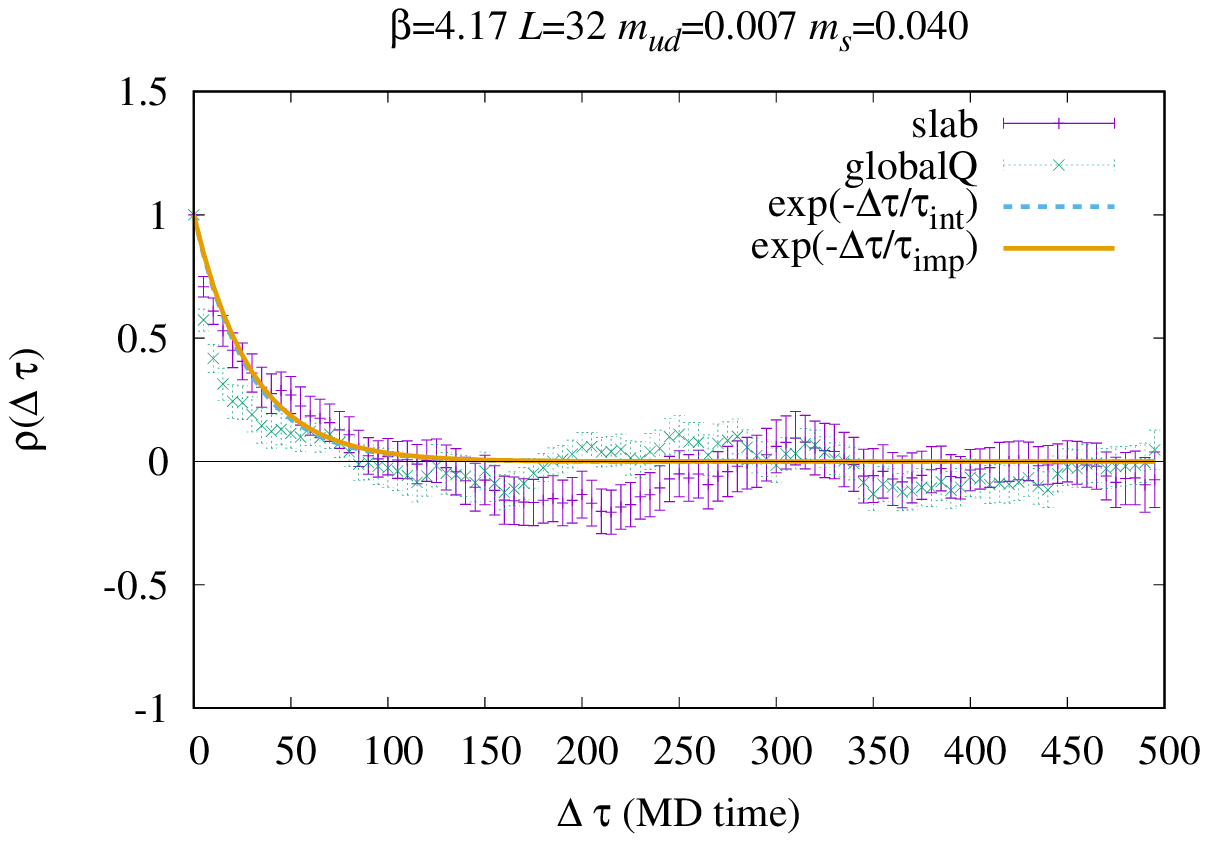}
 \includegraphics[width=8cm]{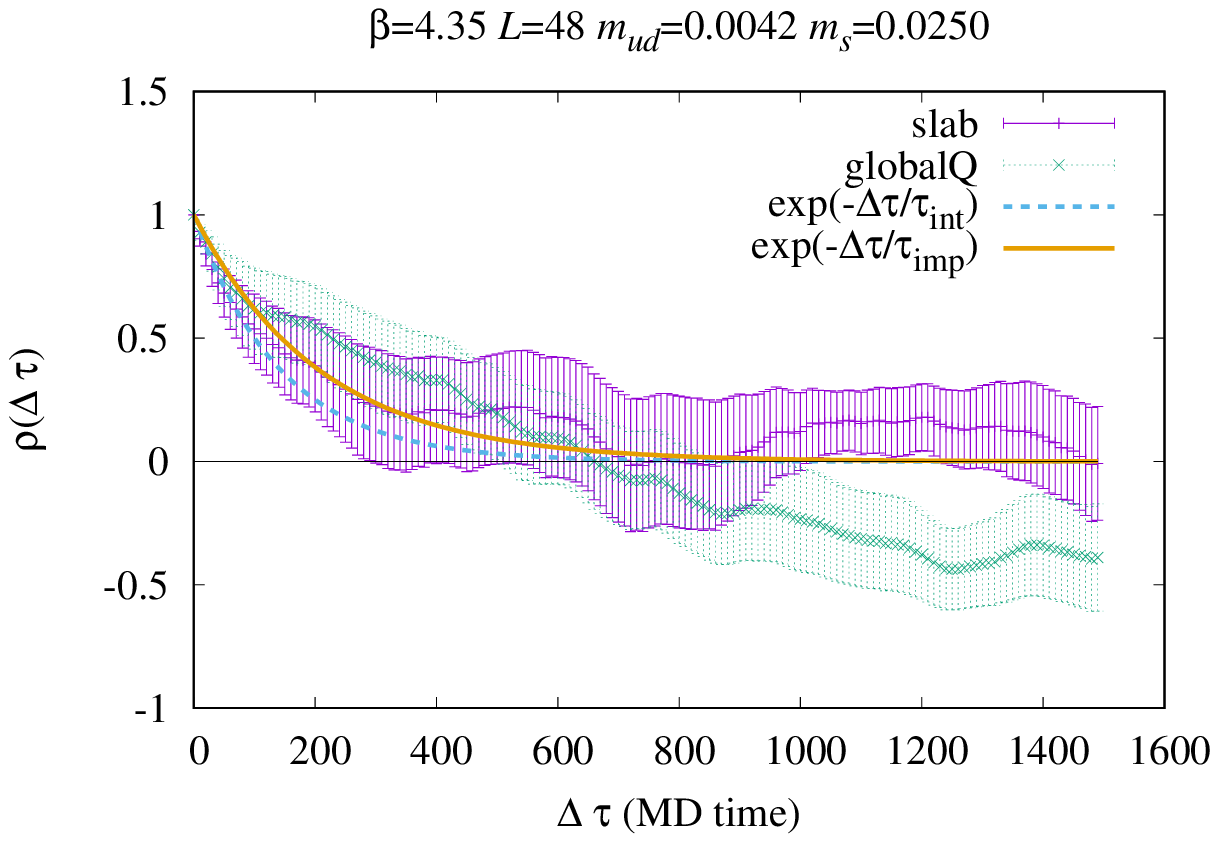}
  \includegraphics[width=8cm]{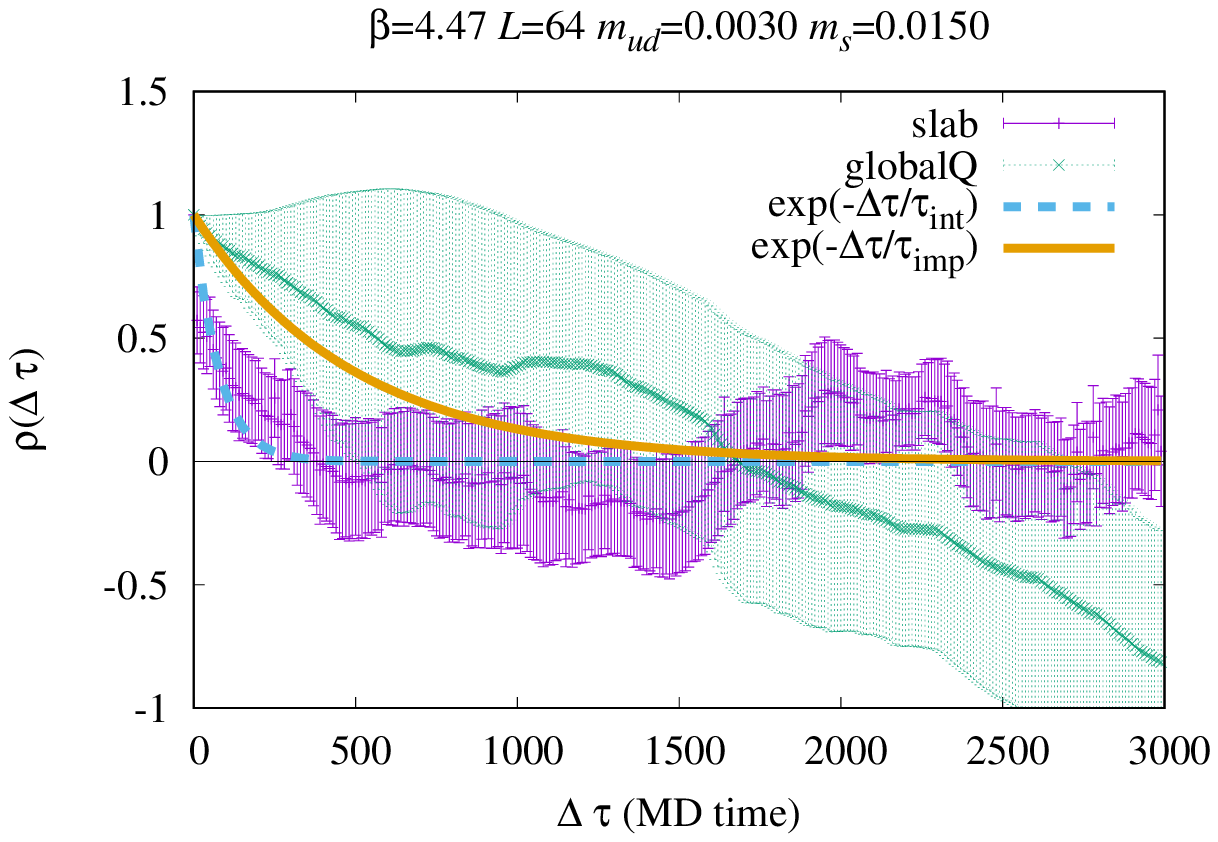}
\caption{
  Auto-correlation function $\rho(\Delta \tau)$ of $\chi_t^{\rm slab}$ (pluses)
  and  $Q_{\rm lat}^2$ (crosses)
  at three different $\beta$
  with a similar pion mass $M_\pi\sim 300$ MeV.
}
\label{fig:autocorrfunc}
\end{figure*}

In lattice QCD simulations,
it is natural to assume that $\tau_{\exp}$ is equal
to the auto-correlation of
the global topological charge.
In our simulations, $\tau_{\exp}$ is estimated by
$\tau_{\rm int}(W)$ of $Q^2_{\rm lat}$, except for $\beta=4.47$ where
we choose $\tau_{\exp}=1700$ MD time by hand (and assuming 100\% error for it),
which is a rough order estimate from the first zero-crossing point of $Q_{\rm lat}$.
Then we compute the auto-correlation of our target observable
$\chi_t^{\rm slab}$ by $\tau_{\rm imp}$ to estimate the error.
The results for $\tau_{\rm imp}$, $\tau_{\exp}$
and $\chi_t^{\rm slab}$ are summarized in Table~\ref{tab:results1}
and the auto-correlation function $\rho(\Delta \tau)$ at three different
$\beta$ with a similar pion mass $M_\pi\sim 300$ MeV
is shown in Fig.~\ref{fig:autocorrfunc}.
At the highest $\beta=4.47$, it is clear that $\chi_t^{\rm slab}$
has shorter auto-correlation time than that of the global topological charge.

With the measured improved auto-correlation time $\tau_{\rm imp}$,
we estimate the statistical errors of $\chi_t^{\rm slab}$
by multiplying $\sqrt{2(\tau_{\rm imp}+\Delta \tau_{\rm imp})/\tau_{interval}}$ to the naive error estimates,
where $\Delta \tau_{\rm imp}$ is the standard deviation of $\tau_{\rm imp}$
and $\tau_{interval}$ denotes the interval trajectory between samples.
The results, as well as the systematic error from the choice of
reference points
are listed in the last column of Table \ref{tab:results1}.

\begin{table}[htbp]
  \centering
  \begin{tabular}{cccc|cccc|c}
    \hline\hline
    $\beta$ & $L$ & $m_{ud}$ & $m_s$ & $M_\pi$ & $\sqrt{2}F_\pi$ & $\tau_{\exp}$ & $\tau_{\rm imp}$ & $\chi_t^{\rm slab}$\\
    \hline
    4.17 & 32 & 0.0035 & 0.04 & 0.09369(32) & 0.05320(19) & 17(04) & 25(9) & 0.217(64)(14)$\times 10^{-5}$\\
    &    & 0.007 & 0.04 & 0.12604(26) & 0.05774(15) & 14(03) & 30(9) &  0.400(78)(21)$\times 10^{-5}$\\
    &    & 0.012 & 0.04 & 0.16267(22) & 0.06254(14) & 65(40) & 62(24) & 1.01(32)(46)$\times 10^{-5}$\\
    &    & 0.019 & 0.04 & 0.20329(19) & 0.06788(14) & 65(40) & 56(22) & 1.59(38)(09)$\times 10^{-5}$\\
    &    & 0.007 & 0.03 & 0.12629(26) & 0.05761(15) & 29(07) & 54(22) & 0.56(15)(53)$\times 10^{-5}$\\
    &    & 0.012 & 0.03 & 0.16179(21) & 0.06190(14) & 74(50) & 71(32) & 0.56(19)(22)$\times 10^{-5}$\\ 
    &    & 0.019 & 0.03 & 0.20302(20) & 0.06730(13) & 56(35) & 42(16) & 1.20(24)(23)$\times 10^{-5}$\\
    & 48 & 0.0035 & 0.04& 0.09203(09) & 0.05440(09) & 38(30) & 21(06) & 0.282(34)(42)$\times 10^{-5}$\\
    \hline
    4.35 & 48 & 0.0042 & 0.025 & 0.08299(18) & 0.03926(11) &  243(153) & 208(114) & 0.91(40)(12)$\times 10^{-6}$\\
    &    & 0.0080 & 0.025 & 0.11312(14) & 0.04291(09) & 318(200) & 362(234) & 2.18(98)(48)$\times 10^{-6}$\\
    &    & 0.0120 & 0.025 & 0.13875(14) & 0.04630(08) & 173(142) & 105(52) & 1.21(32)(07)$\times 10^{-6}$\\
    &    & 0.0042 & 0.018 & 0.08219(19) & 0.03901(11) & 111(49) & 158(72) & 0.59(37)(12)$\times 10^{-6}$\\
    &    & 0.0080 & 0.018 & 0.11284(15) & 0.04275(08) &  236(148) & 220(126) & 0.55(53)(19)$\times 10^{-6}$\\
    &    & 0.0120 & 0.018 & 0.13799(13) & 0.04603(09) & 97(43) & 170(82) & 1.70(53)(21)$\times 10^{-6}$\\
    \hline
    4.47 & 64 & 0.0030 & 0.015 & 0.06316(15) & 0.03141(09) & [1700] & 492(836) & 0.20(27)(09)$\times 10^{-6}$\\
    \hline
  \end{tabular}
  \caption{
   Results for the pion mass $M_\pi$, decay constant $\sqrt{2}F_\pi$,
    $\tau_{\rm exp}$, $\tau_{\rm imp}$, and $\chi_t^{\rm slab}$. $\tau_{\rm exp}$ for $\beta=4.47$ is
   estimated from the first zero-crossing point of $Q_{\rm lat}$. All the data are shown in the lattice units.
   For $\chi_t^{\rm slab}$, the first error denotes the statistical error, while the second shows the systematic error
   due to the effect of freezing global topological charge.
\label{tab:results1}}
\end{table}

%%%%%%%%%%%%%%%%%%%%%%%%%%%%%%%%%%%%%%%%%%%%%%%%
\section{Chiral and continuum limit}
\label{sec:finalresults}
%%%%%%%%%%%%%%%%%%%%%%%%%%%%%%%%%%%%%%%%%%%%%%%%

Figure~\ref{fig:result} presents 
our data for $\chi_t^{\rm slab}$ from all ensembles
plotted in physical units.
%The error-bar in the plot represents
%the statistical and systematic errors
%in the choice of the references $t_1$ and $t_2$, added in quadrature.
The horizontal axis, the quark mass defined in the $\overline{\mbox{MS}}$ scheme
at 2 GeV is 
%%%  MSbar quark mass %%%%%%%%%%%
\begin{eqnarray}
  m_{ud}^{\scriptsize \overline{\mbox{MS}}} = (m_{ud}+m_{res})/Z_S,
\end{eqnarray}
where the renormalization factor $Z_S$ is nonperturbatively computed
in \cite{Tomii:2016xiv}: $Z_S=1.037,\;0.934$, and $0.893$
for $\beta=4.17,\;4.35$, and $4.47$, respectively.
In contrast to the results by other groups with non-chiral fermions,
  we find no strong dependence on $\beta$.
%We find a linear suppression near the chiral limit.
%We also find that there is no clear
%dependence on $\beta$ and $m_s$.

\begin{figure*}[bthp]
  \centering
  \includegraphics[width=8cm]{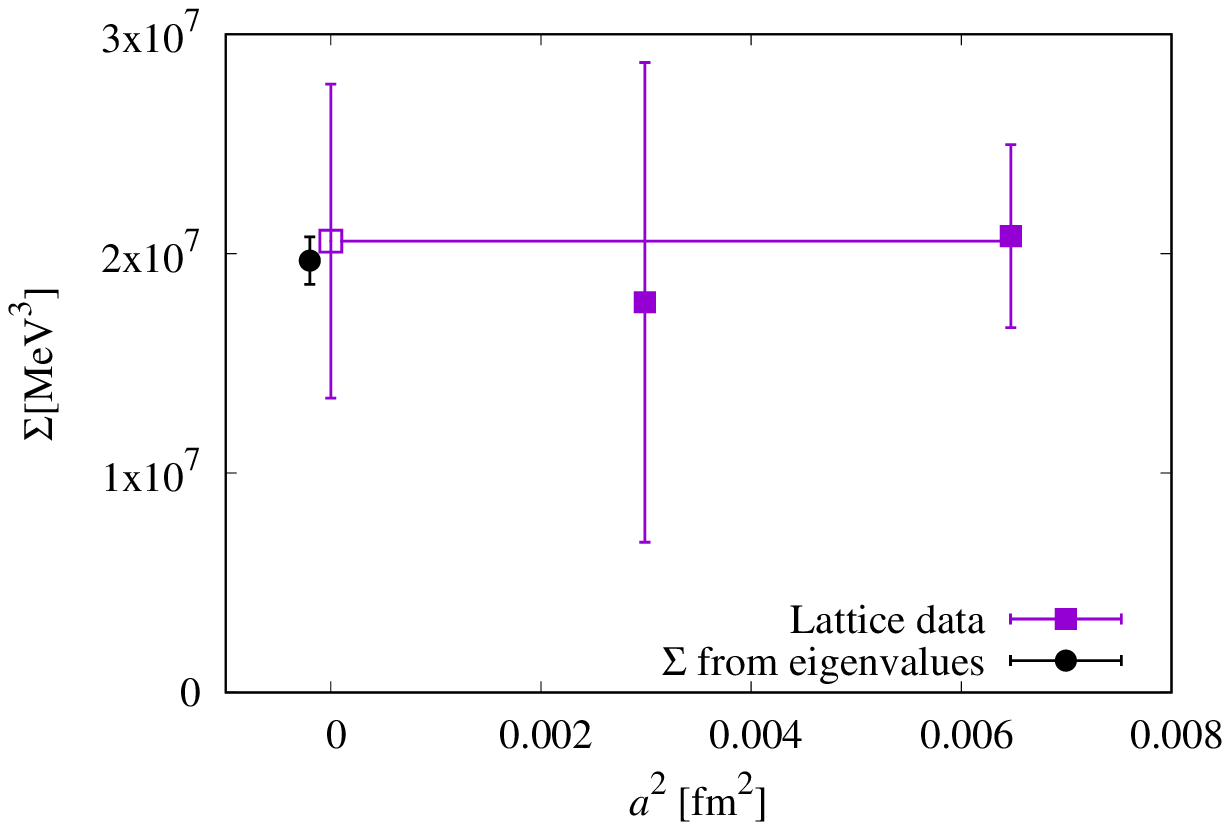}
  \includegraphics[width=8cm]{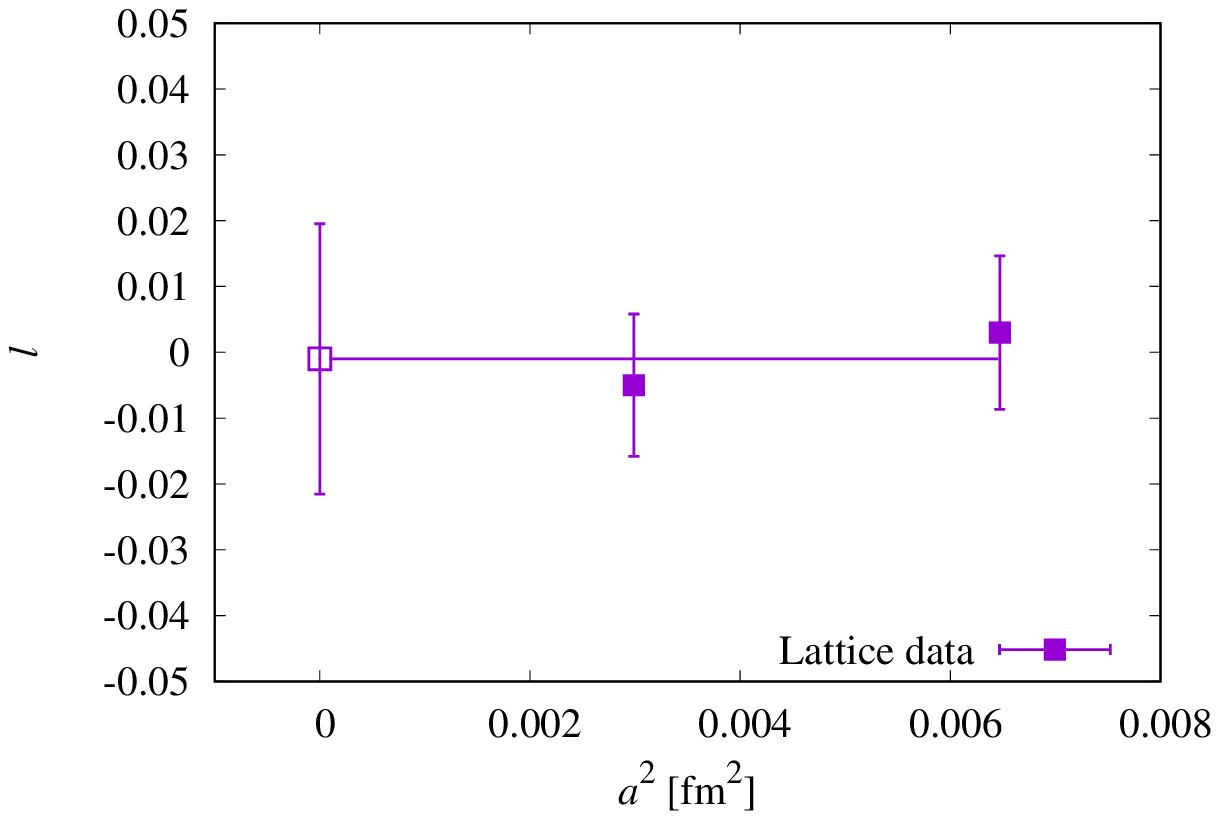}
  \includegraphics[width=8cm]{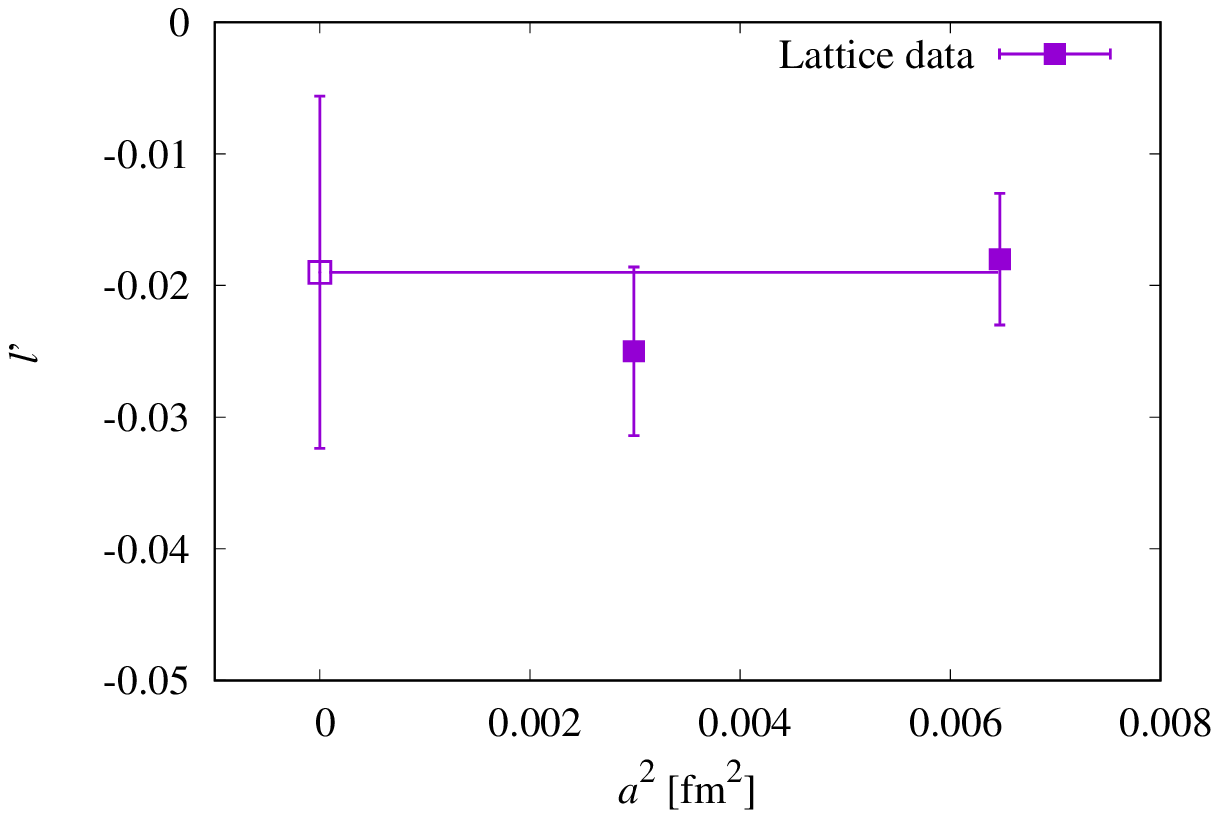}
  \includegraphics[width=8cm]{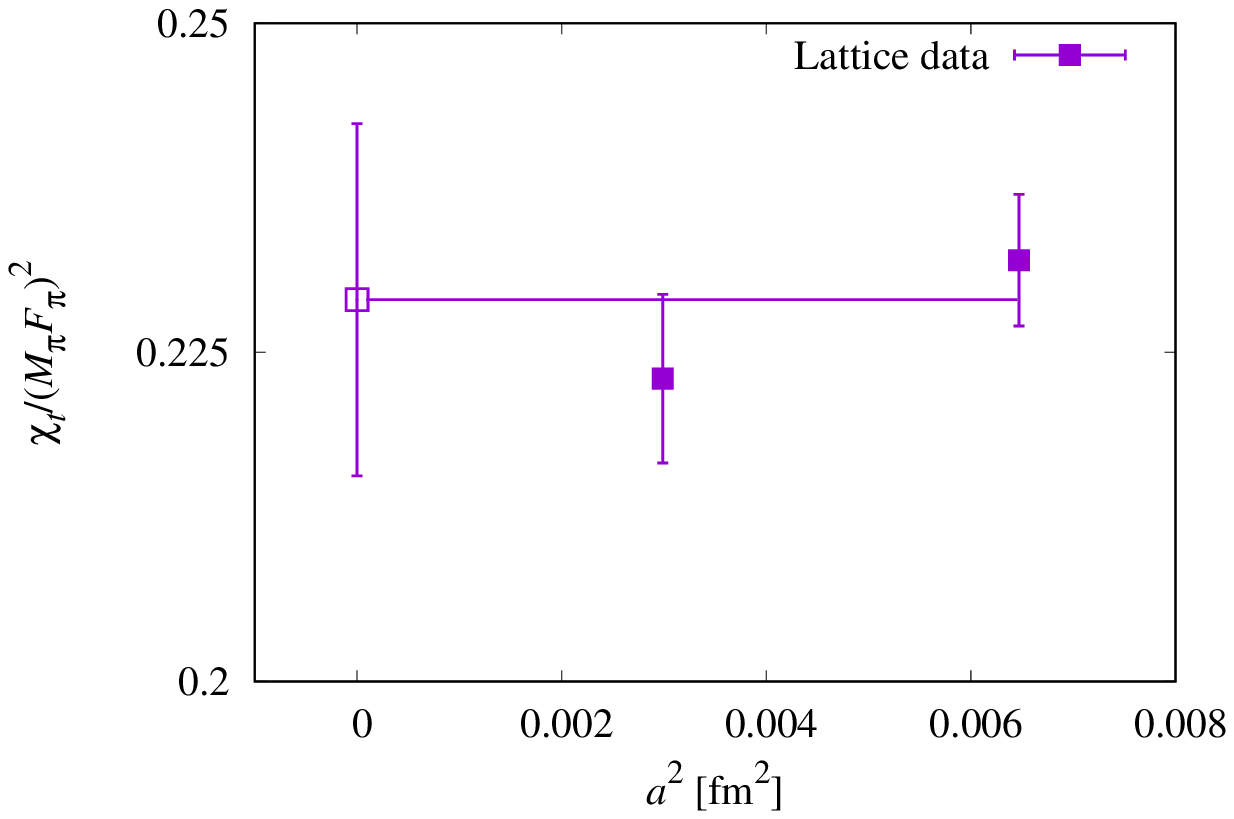}
\caption{
  Continuum limit of $\Sigma$ (we also plot our recent result \cite{Cossu:2016eqs} obtained from the Dirac eigenvalue density),
  $l$, $l^\prime$ and
  $\chi_t^{\rm slab}/(M_\pi F_\pi)^2$ estimated by a constant fit at the physical point.
}
\label{fig:continuumlimit}
\end{figure*}

First, we compare our results directly to the ChPT formula (\ref{eq:ChPT}).
We perform a two-parameter ($\Sigma$ and $l$) fit
to the data at $\beta=4.17$ (solid curve in Fig.~\ref{fig:result})
and $\beta \ge 4.35$ (dashed curve) separately\footnote{
  Since $\beta=4.47$ is simulated at only one choice of the quark masses,
  we simply add the data as one of the $\beta=4.35$ ensembles.
  In fact, $\chi_t$ values at $\beta=4.35$ and 4.47 at the pion mass $\sim 300$ MeV are consistent with each other.
}.
The results for $\Sigma$ and $l$ are listed in Table~\ref{tab:results2}.
Here we also perform the same fit but omitting the heaviest two points,
and take the difference as an estimate for the systematic error
in the chiral extrapolation.
Since the heaviest points have several problems, 1) a strong bias is seen in the global topology,
2) ChPT is less reliable, and 3) mismatch between different $\beta$,
we take the result without them as our central values.
Note, however, this inclusion/elimination affects $l$ but $\Sigma$ is stable against
the change in the fit-range.
Namely, the chiral condensate $\Sigma$ is determined by the low quark mass data.
We then estimate the continuum limit by a constant fit,
as shown in the top two panels in Fig.~\ref{fig:continuumlimit}.
Comparing our result from the constant fit with the linear extrapolation of the central values,
we take the difference as an estimate for the systematic error
in the continuum limit.
In the plots in Fig.~\ref{fig:continuumlimit}, all these errors are
added in quadrature.

Next, using our data for the pion mass $M_\pi$ and decay constant $F_\pi$
together with $\chi_t^{\rm slab}$, 
obtained from each ensemble, we take the ratio given in ~(\ref{eq:ratio}).
By a linear one-parameter fit, we determine $l^\prime$ and
the ratio $\chi_t^{\rm slab}/(M_\pi F_\pi)^2$ at the physical point.
In the same way as the determination of $\Sigma$ and $l$,
we take the chiral and continuum limits of both quantities.
Note that the fixed chiral limit at $1/4$ of the ratio helps us 
to determine these quantities.

Finally let us discuss other possible systematic effects.
In our analysis, the ensembles satisfying $M_\pi L>3.9$ are used 
and  we do not expect any sizable finite volume effects.
In particular, our lightest mass point has $M_\pi L=4.4$.
%The small violation of the chiral symmetry may give an additive
%contribution to ~(\ref{eq:ChPT}).
%However, we find that adding a constant as a free parameter in the fit
%gives a tiny contribution consistent with zero: 5.3(7.8)$\times 10^7$ [MeV$^4$].
We have used configurations at the YM gradient flow--time
around $\sqrt{8t}\sim 0.5$ fm.
We confirm that the flow--time dependence is negligible in the range
$0.25$ fm $< \sqrt{8t} < 0.5$ fm.
c% as shown in Fig.~\ref{fig:tdep}.
We conclude that all these systematic effects are 
negligibly small compared to the statistical and systematic errors given above.

\begin{table}[htbp]
  \centering
  \begin{tabular}{c|cccc}
    \hline\hline
    $\beta$ & $\Sigma^{1/3}$(MeV) & $l$ & $l^\prime$ & $\frac{\chi_t^{\rm slab}}{M_\pi^2 F_\pi^2}$ at physical point \\
    \hline
    4.17 & 275(13)(13) & 0.003(06)(10) & $-$0.018(03)(03) & 0.232(04)(03)\\
    $\geq$ 4.35 & 261(50)(19) & -0.005(09)(06) & $-$0.025(05)(04) & 0.223(05)(04)\\
    \hline
    continuum limit & 274(13)(25)(15)& -0.001(05)(06)(19) & $-$0.019(03)(01)(13) & 0.229(03)(01)(13)\\ 
    \hline
  \end{tabular}
  \caption{
    Our results for $\Sigma$, $l$, $l^\prime$, and $\chi_t^{\rm slab}/(M_\pi F_\pi)^2$ at the physical point.
    The first error denotes the the statistic fluctuation at each simulation point,
    including the effect of long autocorrelation of global topology.    
    The second is the systematic error in chiral extrapolation,
    and the third error denotes that in the continuum limit estimates. See the main text for the details.
\label{tab:results2}}
\end{table}

\begin{figure*}[bthp]
  \centering
 \includegraphics[width=12cm]{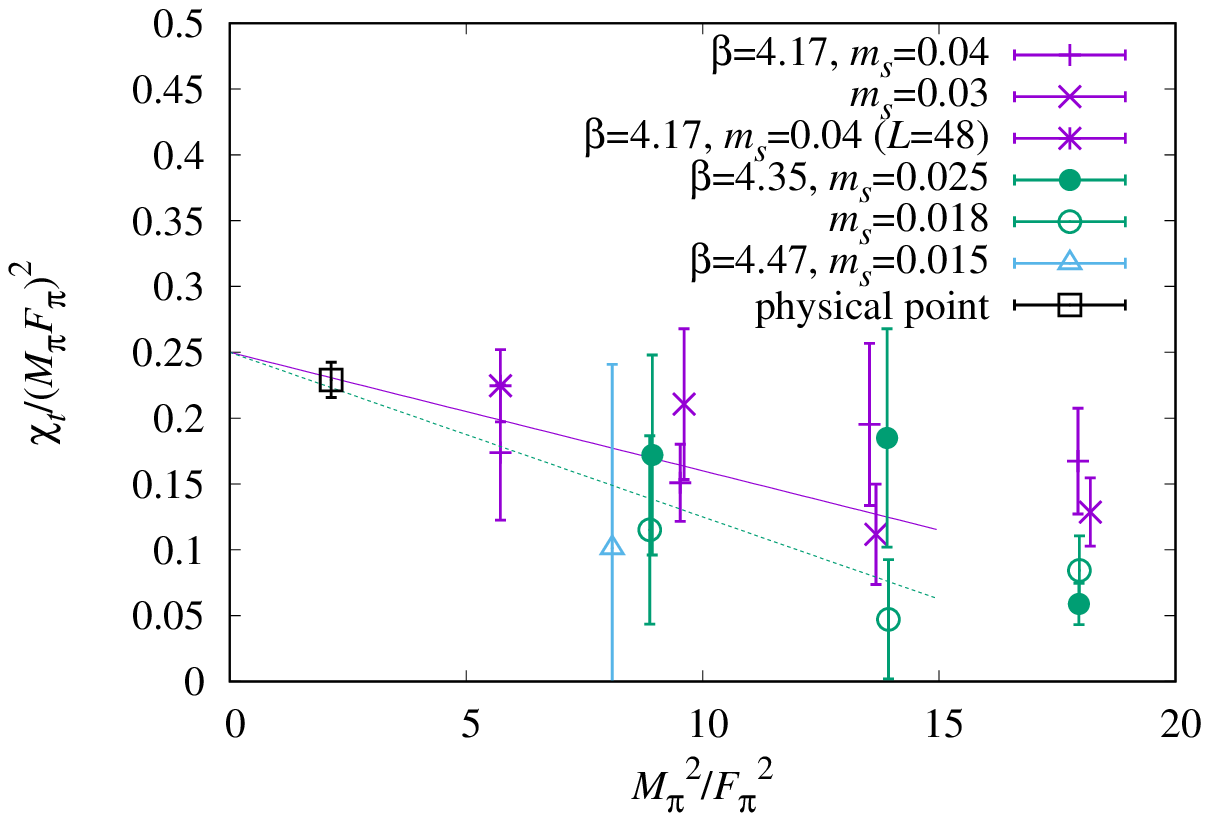}
\caption{
  $m_{ud}$ dependence of the ratio $\frac{\chi_t}{M_\pi^2F_\pi^2}$.
  The data at $M_\pi^2/F_\pi^2>15$ are not included in the fit.
}
\label{fig:ratio}
\end{figure*}

%%%%%%%%%%%%%%%%%%%%%%%%%%%%%%%%%%%%%%%%%%%%%%%%
\section{Summary}
\label{sec:summary}
%%%%%%%%%%%%%%%%%%%%%%%%%%%%%%%%%%%%%%%%%%%%%%%%

% renormalization
With dynamical M\"obius domain wall fermions and the new method
using sub-volume of the simulated lattice,
we have computed the topological susceptibility of QCD.
Its quark mass dependence is consistent with
the ChPT prediction, from which we have obtained
%%% Eq final results
\begin{eqnarray}
\chi_t &=& 0.229(03)(01)(13) M_\pi^2 F_\pi^2\;(\mbox{at physical point}),\\
\Sigma^{\overline{\rm MS}}(\mbox{2GeV}) &=& [274(13)(25)(15)\mbox{MeV}]^3,
\end{eqnarray}
where the first error comes from the statistical uncertainty
at each simulation point, including the effect of freezing topology.
The second and third represent the systematics in the chiral and continuum limits, respectively.
The value of $\Sigma$ is consistent with our recent determination through Dirac spectrum \cite{Cossu:2016eqs}.
We have also estimated the NLO coefficient
%%%  lbar 
\begin{eqnarray}
l &=&(l_3^r-l_7^r+h_1^r-h_3^r)=-0.001(05)(06)(19),\\
l^\prime &=& (-l_4^r-l_7^r+h_1^r-h_3^r)= -0.019(03)(01)(13),
\end{eqnarray}
where $l$ is renormalized at the physical pion mass, while $l^\prime$ is renormalization invariant.
%%% consistent with http://arxiv.org/pdf/1209.4367..    http://arxiv.org/abs/1203.0508
It is interesting to note that $l$ and $l^\prime$ include
a combination of the coefficients $h_1^r-h_3^r$, 
which is supposed to be {\it unphysical}
in ChPT unless $\theta$ dependence is considered.
These are important for possible
couplings of QCD to axions \cite{diCortona:2015ldu}.

\vspace*{5mm}
%%% Result

\vspace*{5mm}
We thank T.~Izubuchi, and other 
members of JLQCD collaboration for fruitful discussions.
We also thank the Yukawa Institute for Theoretical Physics, 
Kyoto University. Discussions during the YITP workshop YITP-T-14-03 
on ``Hadrons and Hadron Interactions in QCD'' were useful in completing this work.
Numerical simulations are performed on IBM System Blue Gene Solution at KEK under 
a support of its Large Scale Simulation Program (No. 16/17-14). 
This work is supported in part 
by the Japanese Grant-in-Aid for Scientific Research
(Nos. JP25800147, JP26247043, JP26400259, JP16H03978), 
%JP25287046, JP25800147, JP26247043, JP26400259, 15K05065
and by MEXT as ``Priority Issue on Post-K computer''
(Elucidation of the Fundamental Laws and Evolution of
the Universe) and by Joint Institute for Computational Fundamental Science (JICFuS).
The work of GC is supported by STFC, grant ST/L000458/1.

\appendix

\section{Effect of strange sea quark}
\label{app:3flavorChPT}

In this work, we have assumed that effect of the  strange quark
is negligible and used $SU(2)$ ChPT
in our main analysis to obtain the chiral extrapolation
of the topological susceptibility.
In this appendix, we consider  $SU(3)$ ChPT and compute possible
correction from the strange quark loop.
We will show that the chiral limit of the ratio (\ref{eq:ratio})
is unchanged even in $SU(3)$ ChPT, which is also protected from finite volume corrections.

The one-loop computation of the topological susceptibility
in general $N_f$-flavor  ChPT was given in \cite{Mao:2009sy, Aoki:2009mx}
and the formula for $N_f=3$ is
%%%  chit SU(3) %%%%%%%%%%%%%%%%%%%%%%%%
\begin{eqnarray}
  \chi_t &=& \bar{m}\Sigma \left[1+\frac{1}{F_\pi^2}\left\{
    -3\frac{\bar{m}}{m_{ud}}\Delta(M_\pi^2)
    -2\left(\frac{\bar{m}}{m_{ud}}+\frac{\bar{m}}{m_{s}}\right)\Delta(M_K^2)
    -\frac{1}{3}\left(\frac{\bar{m}}{m_{ud}}+\frac{2\bar{m}}{m_{s}}\right)\Delta(M_\eta^2)\right.\right.
    \nonumber\\
&&\left.\left.    +16L^r_6(2M_\pi^2+M^2_{ss})+48(3L_7+L^r_8)\bar{M}^2
    \right\}\right],
\end{eqnarray}
where $\bar{m}=m_{ud}m_s/(2m_s+m_{ud})$, 
$M_\pi$,$M_K$, and $M_\eta$ are the (simulated) pion, kaon and $\eta$ meson masses,
respectively. We have also used notations for $M^2_{ss}=2m_s\Sigma/F_\pi^2$, and
$\bar{M}^2=2\bar{m}\Sigma/F_\pi^2$.
The chiral logarithm is expressed by
%%% Eq Delta definition %%%%%%%%%%%%%%%%%%%%%
\begin{eqnarray}
\Delta(M^2) = \frac{M^2}{16\pi^2}\ln \frac{M^2}{\mu_{sub}^2}+g_1(M^2),
\end{eqnarray}
where $\mu_{sub}$ denotes the renormalization scale,
and $g_1$ is finite volume correction (see \cite{Aoki:2009mx} for the details).
In the above formula, we can see three NLO low-energy constants \cite{Gasser:1984gg}:
$L_6^r$ and $L_8^r$ are those renormalized at $\mu_{sub}$,
while $L_7$ is a renormalization scheme independent constant.

One-loop corrections to the pion mass and decay constant were computed
in \cite{Gasser:1984gg}:
%%% Eq pion mass and decay const. %%%%%%%%%%%%
\begin{eqnarray}
  M_\pi^2 = M^2\left[1-\frac{1}{F_\pi^2}\left\{
    -\frac{1}{2}\Delta(M_\pi^2)+\frac{1}{6}\Delta(M_\eta^2)
    +8(L_4^r-2L_6^r)(2M_\pi^2+M_{ss}^2)
    +8(L_5^r-2L_8^r)M_\pi^2
    \right\}\right],\nonumber\\
\end{eqnarray}
and
%%% Eq pion mass and decay const. %%%%%%%%%%%%
\begin{eqnarray}
  F_\pi^2 = F^2\left[1-\frac{1}{F_\pi^2}\left\{
    2\Delta(M_\pi^2)+\Delta(M_K^2)
    -8L_4^r(2M_\pi^2+M_{ss}^2)
    -8L_5^rM_\pi^2
    \right\}\right],
\end{eqnarray}
where $M$ and $F$ are the tree-level mass and decay constant, respectively.

Now let us take the ratio of $\chi_t$ and $M_\pi^2F_\pi^2$.
Noting
%%% Eq mbar expansion %%%%%%%%%%%%%%%%%%%%%%%%
\begin{eqnarray}
  \bar{m}\sim \frac{m}{2}\left(1-\frac{m_{ud}}{2m_s}\right)
  \sim \frac{m}{2}\left(1-\frac{M_\pi^2}{2 M_{ss}^2}\right),
\end{eqnarray}
we obtain
%%% Eq ratio Nf=3 %%%%%%%%%%%%%%%%%%%%%%%%%%%%%%%%%%%%%%
\begin{eqnarray}
  \frac{\chi_t}{M_\pi^2F_\pi^2}&=& \frac{1}{4}\left[1+\frac{2 M_\pi^2 l^{\prime }_{({\rm eff})}}{F_\pi^2} + \mathcal{O}(M_\pi^4)\right],
\end{eqnarray}
where both of strange quark effect, as well as finite volume effects from one-loop diagrams are absorbed in
the (re)definition of
%%% Eq L_7 eff %%%%%%%%%%%%%%%%
\begin{eqnarray}
  l^{\prime}_{({\rm eff})} = -\frac{1}{4M_{ss}^2}\left(F_\pi^2+\Delta(M_K^2)+\frac{1}{2}\Delta(M_\eta^2)\right)+36 L_7 +4L_8^r.
\end{eqnarray}
We, therefore, conclude that the one-loop formula (\ref{eq:ratio})
is valid even when strange quark gives nontrivial effect,
and is also stable against possible finite volume corrections.
This observation helps us in determining $\chi_t$ at the physical point.

\section{Bias from global topology}
\label{app:bias}

In this appendix, we discuss systematics
due to freezing of the global topological charge.
Combining the formulas in \cite{Bietenholz:2015rsa} and \cite{Aoki:2007ka},
the slab topological charge squared at fixed topology of $Q$ becomes
%%% Eq. slab formula fixed topology %%%%%%%
\begin{eqnarray}
\label{eq:slabfixedQ}
\langle Q_{\rm slab}^2(T_{\rm cut}) \rangle_Q = (\chi_t V)\times \frac{T_{\rm cut}}{T}
+\frac{T_{\rm cut}^2}{T^2}\left(Q^2-\chi_t V\right)
+C,
\end{eqnarray}
for $0\ll T_{\rm cut}\ll T$.
Therefore, if the global topological charge $Q$ were badly sampled and
its average of $Q^2$ in the ensemble deviated from $\chi_tV$,
we should have a quadratic term in $T_{\rm cut}$ as
%%% Eq. slab formula fixed topology %%%%%%%
\begin{eqnarray}
\label{eq:slabfixedQ}
\langle Q_{\rm slab}^2(T_{\rm cut}) \rangle_{biased}
%&=& (\chi_t V)\frac{T_{\rm cut}}{T}+\frac{T_{\rm cut}^2}{T^2}\left(\langle Q^2\rangle_{biased}-\chi_t V\right)
%+C\nonumber\\
&=&
(\chi_t V)\frac{T_{\rm cut}}{T}
\left[1+ \frac{T_{\rm cut}}{\chi_t VT}\left(\langle Q^2\rangle_{biased}-\langle Q\rangle^2_{biased}-\chi_t V\right)\right]+C,
%\nonumber\\&&
%+C(1-e^{-m_0 T_{\rm cut}})(1-e^{-m_0 (T-T_{\rm cut})}),
\end{eqnarray}
where $\langle \cdots \rangle_{biased}$ denotes the estimate
obtained from a biased sampling of configurations.
Here we have included the term $\langle Q\rangle^2_{biased}$, which comes from
the use of the subtracted operator $q^{\rm lat}-\langle Q/V\rangle_{biased}$ in our numerical analysis.

If the correction  $\frac{T_{\rm cut}}{(\chi_t V)T}\left(\langle Q^2\rangle_{biased}-\langle Q\rangle^2_{biased}-\chi_t V\right)$
is small, our original linear $+$ constant formula is still valid.
As the correction is proportional to $\frac{T_{\rm cut}}{T}$,
if we have a window $T_{\rm cut}\ll T$, or the freezing $\langle Q^2\rangle_{biased}-\langle Q\rangle^2_{biased}$
happens to be near the true value of $\chi_tV$, we can still extract $\chi_t$ from the linear slope
(this seems to happen on the data at $\beta=4.47$).

In order to estimate the systematics due to the correction term,
we compare the results 
with 1) those obtained from different reference times $(t_1', t_2')=(t_1,\frac{t_1+t_2}{2})$,
and $(\frac{t_1+t_2}{2},t_2)$\footnote{
This also tests if the effect from the excited state $m_0$ is small or not.
}, 
and 2) those obtained without the subtraction of $\langle Q\rangle/V$ in the definition of the topological charge density.
Then we take the larger deviation as the systematic error.
Since a part of $\langle Q\rangle^2_{biased}$ is expected to be canceled by $\langle Q^2\rangle_{biased}$,
this analysis is rather conservative.
As presented in Tab.~\ref{tab:results1}, the deviations are
comparable to the statistical errors.

Let us look into our ``worst'' case, the data at $\beta=4.35$ and $(m_{ud},m_s)=(0.012,0.018)$ in our ensembles,
which shows the strongest curvature.
As is expected, the global topological charge sampling is biased:
the estimate for $\langle Q^2\rangle = 12(4)$ in the
former half (0-2500 MD time)
of the simulation time, is quite different from $\langle Q^2\rangle = 40(17)$
in the latter half (2500-5000 MD time).
But the obtained values of $\chi_t^{\rm slab}$ show a milder deviation,
$1.30(53)\times 10^{-6}$ for the former half and $1.89(64)\times 10^{-6}$ for the latter,
which are consistent within errors.
This analysis\footnote{
  We thank W. Bietenholz and P. de Forcrand for suggesting this analysis of freezing topology effects.
} shows that the systematics due to freezing topology is under control,
at least, at the level of the statistical errors.
Our ChPT fit with reasonable $\chi^2/d.o.f.\sim 1.4$
also supports our conclusion.

\end{document}